\documentclass[twocolumn]{aastex63}
\usepackage[]{units}
\usepackage[update,prepend]{epstopdf}
\usepackage{graphicx}
\usepackage{amssymb}
\usepackage{amsmath}
\usepackage{threeparttable}
\usepackage{hyperref}
\usepackage{color}

\usepackage{mathtools}

\usepackage{makecell}

\newcommand{\kms} {\,km\,s$^{-1}$}
\newcommand{\masyr} {\,mas\,yr$^{-1}$}

\newcommand{\Msun}{\,M$_\odot$}

\mathchardef\mhyphen="2D
\shorttitle{Eccentricity of wide binaries}
\shortauthors{Hwang et al.}

\begin{document}

\title{The eccentricity distribution of wide binaries and their individual measurements}

\correspondingauthor{Hsiang-Chih Hwang}
\email{hchwang@ias.edu}

\author[0000-0003-4250-4437]{Hsiang-Chih Hwang}
\affiliation{Institute for Advanced Study, Princeton, 1 Einstein Drive, NJ 08540, USA}
\affiliation{Department of Physics \& Astronomy, Johns Hopkins University, Baltimore, MD 21218, USA}

\author[0000-0001-5082-9536]{Yuan-Sen Ting}
\affiliation{Research School of Astronomy \& Astrophysics, Australian National University, Cotter Rd., Weston, ACT 2611, Australia}
\affiliation{Research School of Computer Science, Australian National University, Acton ACT 2601, Australia}

\author[0000-0001-6100-6869]{Nadia L. Zakamska}
\affiliation{Institute for Advanced Study, Princeton, 1 Einstein Drive, NJ 08540, USA}
\affiliation{Department of Physics \& Astronomy, Johns Hopkins University, Baltimore, MD 21218, USA}

\begin{abstract}

Eccentricity of wide binaries is difficult to measure due to their long orbital periods. With {\it Gaia}'s high-precision astrometric measurements, eccentricity of a wide binary can be constrained by the angle between the separation vector and the relative velocity vector (the $v$-$r$ angle). In this paper, by using the $v$-$r$ angles of wide binaries in {\it Gaia} Early Data Release 3, we develop a Bayesian approach to measure the eccentricity distribution as a function of binary separations. Furthermore, we infer the eccentricities of individual wide binaries and make them publicly available. Our results show that the eccentricity distribution of wide binaries at $10^2$\,AU is close to uniform and becomes superthermal at $>10^{3}$\,AU, suggesting two formation mechanisms dominating at different separation regimes. The close binary formation, most likely disk fragmentation, results in a uniform eccentricity distribution at $<10^{2}$\,AU. The wide binary formation that leads to highly eccentric wide binaries at $>10^{3}$\,AU may be turbulent fragmentation and/or the dynamical unfolding of compact triples. With {\it Gaia}, measuring eccentricities is now possible for a large number of wide binaries, opening a new window to understanding binary formation and evolution.

\end{abstract}

\section{Introduction}

Eccentricity is one of the fundamental orbital parameters in orbital dynamics. Eccentricity provides key constraints on the binary formation mechanisms \citep{Duquennoy1991,Duchene2013}. In the hierarchical three-body systems, the excitation of inner orbit's eccentricity through the Kozai-Lidov mechanism \citep{Kozai1962, Lidov1962} is one important formation channel for close binaries \citep{Kiseleva1998,Eggleton2001,Fabrycky2007} and hot jupiters \citep{Fabrycky2007,Dawson2018}. If the outer orbit is eccentric, the secular evolution of triples is chaotic and can further enhance binary mergers and result in exotic systems like retrograde hot jupiters \citep{Naoz2011,Naoz2013,Naoz2014,Naoz2016}.

Eccentricity measurement is challenging for resolved binaries with separations $>100$\,AU due to their long orbital periods. For example, an equal-solar-mass binary at $100$\,AU has an orbital period of 700\,yr, and typical monitoring observations with a few years' baseline will only reveal a tiny fraction of the orbit and not result in an orbital solution. Therefore, despite of its importance for binary formation and three-body interaction, the eccentricity of wide binaries remains poorly quantified.  

\begin{figure}
	\centering
	\includegraphics[width=\linewidth]{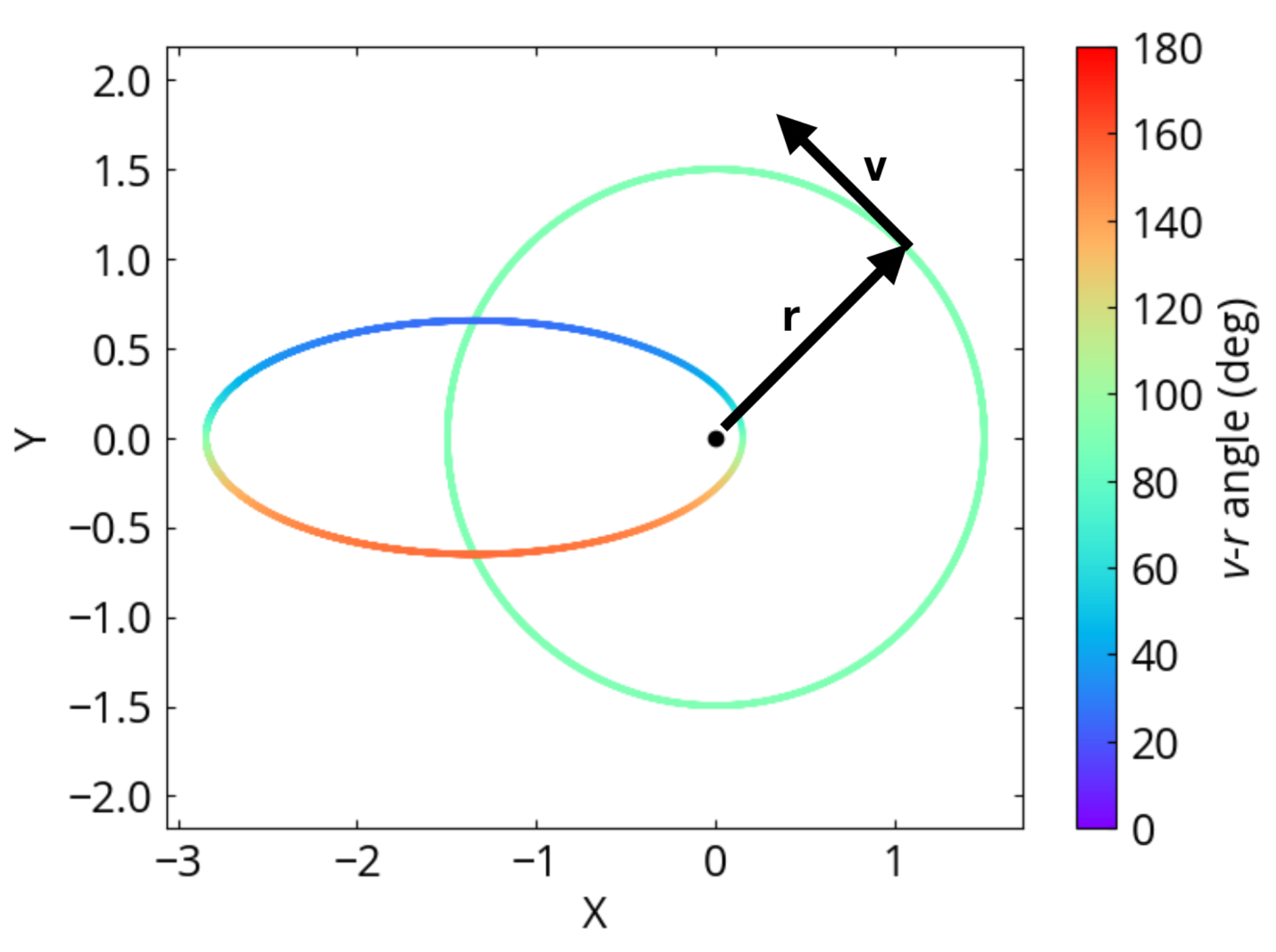}
	\caption{A face-on view of a circular and an eccentric orbit ($e=0.9$), with colors indicating the $v$-$r$ angles, the angle between the separation vector (`r' arrow) and the relative velocity vector (`v' arrow). The black point is the focus of the orbit. The horizontal and vertical axes are physical distances with arbitrary units for the relative separation vector $\vec{r}=\vec{r}_1-\vec{r}_0$. Therefore, $v$-$r$ angles provide critical information about orbital eccentricities. }
	\label{fig:orbit}
\end{figure}

There is one particular observable in wide binaries that is tightly related to the orbital eccentricity: the angle between the separation vector and the relative velocity vector, dubbed $v$-$r$ angle. The $v$-$r$ angle of a face-on circular orbit is always 90$^\circ$, and that of a face-on eccentric orbit is not 90$^\circ$\ in general, depending on the exact eccentricity and orbital phase (Fig.~\ref{fig:orbit}). Therefore, once the orbital phase and viewing angle are taken into account, the $v$-$r$ angle provides an opportunity for inferring the eccentricity (distribution) of wide binaries without observationally expensive monitoring.

The concept of using $v$-$r$ angles to measure the eccentricities was proposed by \cite{Tokovinin1998}, who demonstrates that different eccentricities result in various distributions of $v$-$r$ angles. Based on a similar idea, \cite{Shatsky2001} shows that wide companions around multiple stars tend to have moderate eccentricities. \cite{Tokovinin2016} further expand the method to include the information on orbital velocities and use the $v$-$r$ angle-orbital velocity space to infer the eccentricities. While the inclusion of relative velocity provides more information to constrain the eccentricity than the $v$-$r$ angle alone, it requires the masses and distances of the binaries, which are not typically accessible for most systems.

The space astrometry mission {\it Gaia}  has revolutionized wide binary research. With high-quality parallaxes and proper motions available for billions of stars, a large sample of wide binaries has been made possible \citep{Oh2017,El-Badry2018b,Hartman2020,Tian2020,El-Badry2021}, resulting in interesting new findings about binary formation and evolution \citep[e.g.][]{El-Badry2019a,Hawkins2020,Hwang2020c}. 

Amazingly, {\it Gaia}'s proper motion precision is sufficient to measure the relative velocity of wide binaries and therefore the $v$-$r$ angle. With these $v$-$r$ angle measurements and the magnitude of relative velocities, \cite{Tokovinin2020a} reports that wide binaries at about $100$\,AU have less eccentric orbits than those at $>1000$\,AU. For binaries at $>1000$\,AU, their eccentricity distribution is close to thermal ($f_e(e)de=2ede$), a theoretical distribution when a population of binaries reaches an equilibrium state \citep{Jeans1919,Ambartsumian1937,Heggie1975,Kroupa2008}.

In this paper, with about one million wide binaries available from {\it Gaia} Early Data Release 3 \citep{El-Badry2021}, we develop a Bayesian scheme to derive the eccentricity distribution of wide binaries and to infer the eccentricity for individual wide binaries. In contrast to the method of \cite{Tokovinin2020a}, our method only uses the $v$-$r$ angles measured from {\it Gaia} without using the magnitude of relative velocity. The advantage of our method is that it does not reply on mass and parallax measurements and therefore can include a dramatically larger number of wide binaries than \cite{Tokovinin2020a}, but this statistical improvement occurs at the expense of larger uncertainties for individual wide binaries' eccentricity measurements.

The structure of this paper is as follows. Section~\ref{sec:sample} explains the relation between eccentricity and the observed $v$-$r$ angle. In Section~\ref{sec:observation}, we present a Bayesian framework to measure the eccentricity distribution as well as to infer the eccentricity for individual binaries. We discuss the implications in Section~\ref{sec:discussion} and conclude in Section~\ref{sec:conclusion}.

\section{Methodology for eccentricity measurements}
\label{sec:sample}

\subsection{Notation for the two-body problem}

We begin with defining the basic notations for the two-body problem. We have a first body of mass $m_0$, position $\vec{r}_0$, velocity $\vec{v}_0=d\vec{r}_0/dt$, and acceleration $\vec{a}_0=d\vec{v}_0/dt$, and a second object of mass $m_1$, position $\vec{r}_1$, velocity $\vec{v}_1=d\vec{r}_1/dt$, and acceleration $\vec{a}_1=d\vec{v}_1/dt$. The origin of the coordinate system is placed at the system's barycenter $\vec{R}=(m_0 \vec{r}_0 + m_1 \vec{r}_1)/m$, where $m:=m_0+m_1$ is the total mass. 

The two-body problem can be described using an effective single-object formulation. The relative separation is $\vec{r} = \vec{r}_1-\vec{r}_0$ and the relative velocity is $\vec{v} = \vec{v}_1-\vec{v}_0$, and their magnitudes are $r=|\vec{r}|$ and $v=|\vec{v}|$. The orbital solution to the two-body problem is a series of conic sections, 
\begin{equation}
\label{eq:1}
r = \frac{a (1-e^2)}{1+e \cos{(\phi - \omega)}},
\end{equation}
where $a$ is the semi-major axis, $e$ is the eccentricity defined between 0 and 1 with $e=0$ being a circular orbit, $\phi$ is the orbital angle that changes with time, and $\omega$ is the longitude of pericenter. The orbital period is $P=2\pi \sqrt{a^3/Gm}$. A running parameter, known as the true anomaly $f$, is defined as $f=\phi - \omega$.

Another running parameter, the eccentric anomaly $u$, is defined by the relations
\begin{equation}
\label{eq:cosu}
\cos u = \frac{\cos f + e}{1 + e \cos f},
\end{equation}
and
\begin{equation}
\label{eq:sinu}
\sin u = \frac{\sqrt{1-e^2} \sin f}{1 + e \cos f}.
\end{equation}
The inverse relations are 
\begin{equation}
\label{eq:cosf}
\cos f = \frac{\cos u - e}{1 - e \cos u},
\end{equation}
and
\begin{equation}
\label{eq:sinf}
\sin f = \frac{\sqrt{1-e^2} \sin u}{1 - e \cos u}.
\end{equation}

Using eccentric anomaly, Eq.~\ref{eq:1} can be rewritten as 
\begin{equation}
\label{eq:eccentric-anomaly}
r = a(1 - e \cos u).
\end{equation}

The true anomaly and eccentric anomaly are useful because they can express Eq.~\ref{eq:1} and Eq.~\ref{eq:eccentric-anomaly} in a simple form analytically, but one drawback is that they are not linear in time in general. For this purpose, another important running parameter for the orbital phase is the mean anomaly $M$, defined as 
\begin{equation}
\label{eq:mean-anomaly}
M\equiv u-e \sin u,
\end{equation}
which is linear in time. Therefore, when we perform a simulation, we uniformly sample the mean anomaly from 0 to 2$\pi$ to have a realistic representation for observations.

\subsection{Viewing angle and the projection effects}
\label{sec:projection-effect}

To specify the Keplerian orbit in the three-dimensional Cartesian coordinate system ($X$, $Y$, $Z$), we include a few additional orbital elements, including the inclination $\iota$, and the longitude of the ascending node $\Omega$. With these orbital parameters, the components of separation vector $\vec{r}=r_X \hat{X} + r_Y \hat{Y} + r_Z \hat{Z}$ and velocity vector $\vec{v} = v_X \hat{X} + v_Y \hat{Y} + v_Z \hat{Z}$ can be derived following \cite{Poisson2014}.

In the 3-dimensional Cartesian coordinate system ($X$, $Y$, $Z$), we place the observer at infinity on the positive-$Z$ axis. At infinity, the observer sees a separation projected on the $X$-$Y$ plane (the sky), $\vec{r}_{XY}=r_X \hat{X} + r_Y \hat{Y}$. The line-of-sight component of the separation vector ($r_Z$) is often difficult to measure observationally as it is limited by the precision of the parallax measurement.

The observer can measure the velocity vector projected on the $X$-$Y$ plane, $\vec{v}_{XY}=v_X \hat{X} + v_Y \hat{Y}$, using the proper motion data. In principle, $\vec{v}_Z$ can be measured if the radial velocities are available for both stars. However, in this paper, we focus on $\vec{v}_{XY}$ but not $\vec{v}_Z$ for two reasons. First, with the {\it Gaia} survey, $\vec{v}_{XY}$ is available for most of the wide binaries, but only $0.3$\% of them have high-precision (errors $<1$\kms) radial velocities for computing $\vec{v}_Z$. Second, the presence of an unresolved close companion can easily affect the radial velocities and $\vec{v}_Z$ at the relevant magnitude because $v\propto a^{-1/2}$, and about half of the wide binaries have unresolved companions \citep{Tokovinin2014b,Moe2017}. Compared $\vec{v}_Z$, $\vec{v}_{XY}$ is less affected by the orbital motion of the unresolved binary, and we discuss it in more detail in Sec.~\ref{sec:systematics}.

When we simulate a sample of binaries, we draw a uniform distribution between 0 and $2\pi$ for the mean anomaly $M$ because $M$ is linear with time. Since the components of $\vec{r}_{XY}$ and $\vec{v}_{XY}$ can be written as a function of true anomaly \citep{Poisson2014}, we derive eccentric anomaly $u$ from $M$ using the iterative Newton's root-finding method to solve Eq.~\ref{eq:mean-anomaly}, and then obtain true anomaly $f$ from $u$ using Eq.~\ref{eq:cosf} and \ref{eq:sinf}. For the case where wide binaries are randomly oriented, their orientations are sampled so that the angular momentum vector is uniform on a sphere.

\subsection{The relation between $v$-$r$ angle and eccentricity}

The angle between $\vec{r}$ and $\vec{v}$, dubbed $v$-$r$ angle or $\gamma_{3D}$, is the primary measure we use in this paper to quantify the eccentricity. Mathematically, the $v$-$r$ angle is 
\begin{equation}
\label{eq:vr-angle-0}
\cos \gamma_{3D} = \frac{\vec{v} \cdot \vec{r}}{v r}.
\end{equation}
The subscript $3D$ indicates that this $v$-$r$ angle $\gamma_{3D}$ is measured using three-dimensional vectors $\vec r$ and $\vec v$, instead of their projected components. Equivalently, $\gamma_{3D}$ is the observed $v$-$r$ angle when the observer views the binary face-on (i.e. the line of sight is perpendicular to the orbital plane). Rephrasing Eq.~\ref{eq:vr-angle-0} in terms of eccentricity yields 
\begin{equation}
\label{eq:vr-angle}
\cos \gamma_{3D} = \frac{e \sin f}{\sqrt{1 + e^2 + 2 e \cos f}},
\end{equation}
where $f$ is the true anomaly. As demonstrated in Fig.~\ref{fig:orbit}, for a circular orbit with $e=0$, $\cos \gamma_{3D}=0$ and thus $\gamma_{3D}=90^\circ$\ regardless of orbital phases, meaning that the separation vector is always perpendicular to the velocity vector. For an eccentric orbit ($e>0$), in general $\gamma_{3D} \ne 90^\circ$ and it varies with the orbital phase. 

The sign of $\cos \gamma_{3D}$ indicates whether the binary is getting closer or coming apart. If $\cos \gamma_{3D}>0$ ($\gamma_{3D} < 90^\circ$), then the binary is currently shortening its binary separation, and vice versa. For Keplerian orbits, the distribution of $\gamma_{3D}$ is symmetric with respect to $\gamma_{3D}=90^\circ$. Unbound binaries that are undergoing disruption and thus expanding their separations have $\gamma_{3D}<90^\circ$, and $\gamma_{3D}$ asymptotes to 0$^\circ$ at late stages of disruption. Therefore, the sign of $\cos \gamma_{3D}$ can potentially carry information about the status of a binary and therefore we do not define $\gamma_{3D}$ only between 0$^\circ$ and 90$^\circ$\ using the symmetry assumption which would be suitable for purely Keplerian orbits. Rather, $\gamma_{3D}$ is defined between 0$^\circ$ and 180$^\circ$\ using Eq.~\ref{eq:vr-angle-0} and we use the symmetry with respect to 90$^\circ$\ as a check to determine whether the binary population is consistent with Keplerian motion or whether there is a disrupted binary population. As a result, we find the observed $v$-$r$ angle distribution symmetric and no significant enhancement at 0$^\circ$, supporting that our sample is dominated by Keplerian orbits.

Eq.~\ref{eq:vr-angle} relates three unitless orbital parameters: the eccentricity ($e$), $v$-$r$ angle ($\gamma_{3D}$), and orbital phase ($f$), without any dependence on the mass and the physical separation of the binary. This is not surprising because all quantities with physical units cancel out on the right-hand side of Eq.~\ref{eq:vr-angle-0}.  

The angle $\gamma_{3D}$ in Eq.~\ref{eq:vr-angle} is not directly observable because the observer only measures the projected separation and velocity. Since we place the observer at infinity on the $+Z$ axis and therefore the $z$-direction is the line-of-sight direction, the observed $v$-$r$ angle, $\gamma$, is 
\begin{equation}
\label{eq:vr-angle-xy}
\cos \gamma = \frac{\vec{v}_{XY} \cdot \vec{r}_{XY}}{v_{XY} r_{XY}},
\end{equation}
where $r_{XY}$ and $v_{XY}$ are the magnitudes of $\vec{r}_{XY}$ and $\vec{v}_{XY}$, respectively. In the rest of the paper, we use the notation $\gamma$ without any subscripts to refer to the observed $v$-$r$ angle measured from the two-dimensional vectors $\vec{r}_{XY}$ and $\vec{v}_{XY}$. By plugging in the components of $\vec{v}_{XY}$ and $\vec{r}_{XY}$, we provide the analytic form of $\cos \gamma$ in Eq.~\ref{eq:gamma-XY-ana}.

Similarly to $\cos \gamma_{3D}$ in Eq.~\ref{eq:vr-angle}, $\cos \gamma$ only depends on unitless orbital parameters in Eq.~\ref{eq:gamma-XY-ana}: eccentricity, orbital phase, and binary orientation (equivalently, the viewing angle). Again, $\cos \gamma$ does not depend on other physical quantities like mass and physical separation. This relation means that, once the orbital phase and the viewing angle are taken into account, we can use the $v$-$r$ angle to infer, probabilistically, the eccentricity for an individual wide binary.

\subsection{Measuring eccentricity from the $v$-$r$ angle}

In this section, we provide the proof of concept for two related ideas. First, we show that it is possible to infer the eccentricity for individual wide binaries using their observed $v$-$r$ angles without the knowledge of the orbital phase and binary orientation. Second, for a sample of wide binaries, we demonstrate that their $v$-$r$ angle distribution is intimately linked to the underlying eccentricity distribution.

There is no one-to-one relation between the $v$-$r$ angle and the eccentricity  because of the unknown orbital phase and the binary orientation. But since both orbital phase and the orientation are random values and their distributions are well understood, we can infer the posterior of eccentricity given the $v$-$r$ angle of a binary.

\begin{figure*}
	\centering
	\includegraphics[height=0.35\linewidth]{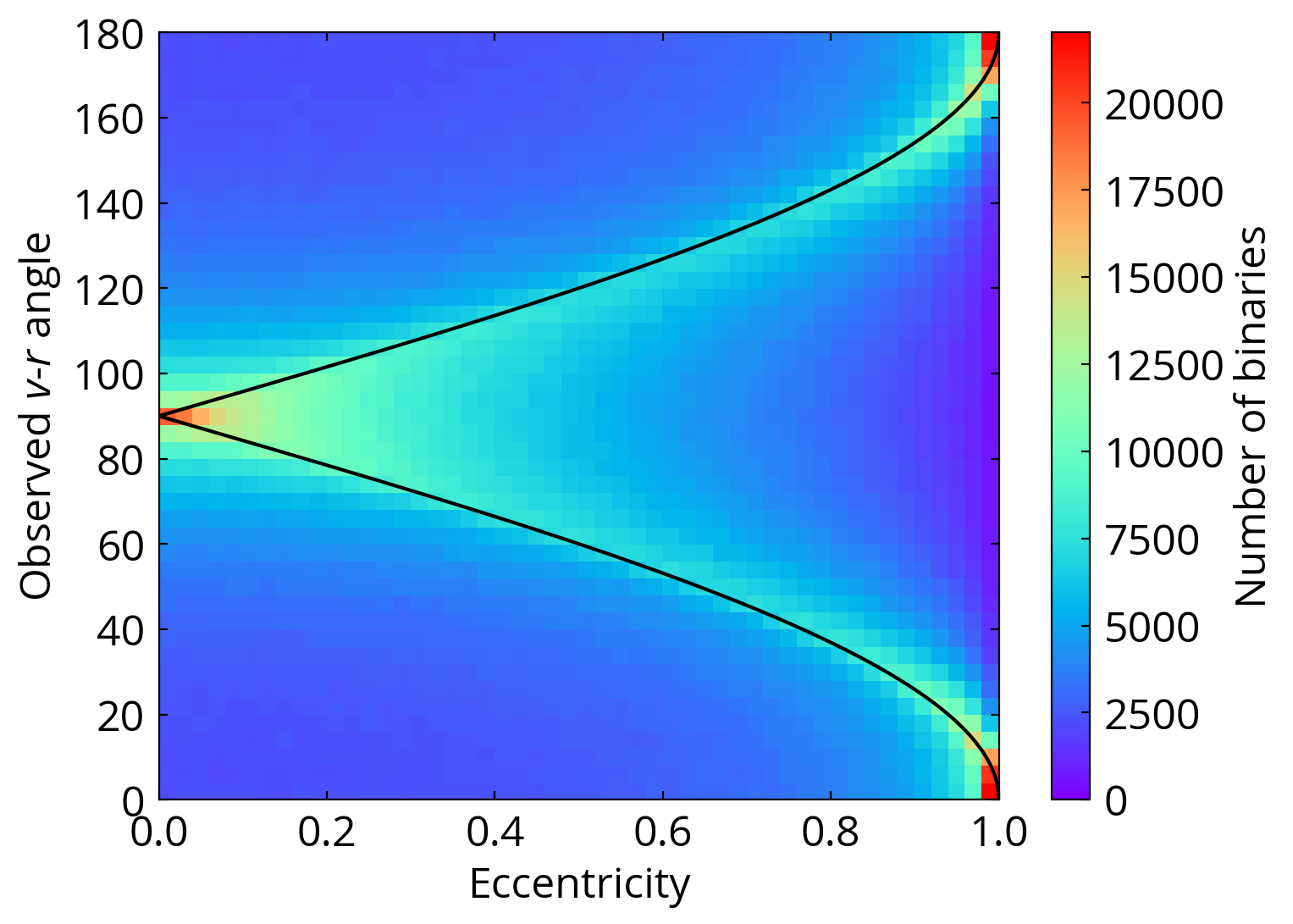}
	\includegraphics[height=0.35\linewidth]{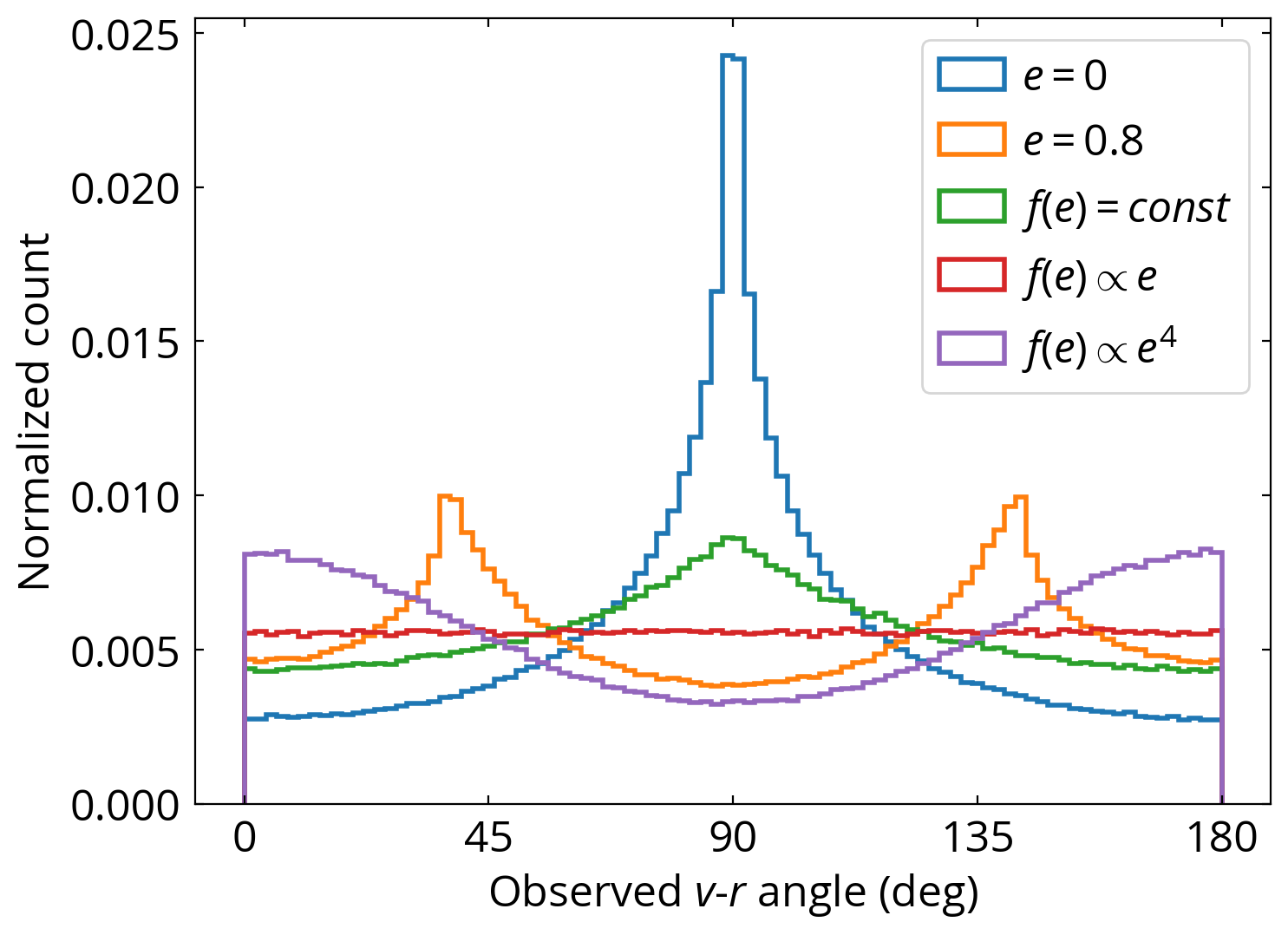}
	\caption{Left: Binary simulations with random binary orientation and a uniform eccentricity distribution. The observed $v$-$r$ angles include the projection effects. The solid black line shows the relation $\left| \cos \gamma \right|=e$, the most probably observed $v$-$r$ angle for a given eccentricity. The observed $v$-$r$ angle is tightly correlated with the eccentricity even when the binary orientation and the orbital phase are not known. Right: Simulated distributions of the observed $v$-$r$ angles for single-valued eccentricities and power-law eccentricity distributions. Different eccentricity distributions have distinctive observed $v$-$r$ angle distributions. }
	\label{fig:vr-e}
\end{figure*}

To demonstrate the connection between eccentricity and $v$-$r$ angle, we simulate a sample of binaries and show their eccentricities and observed $v$-$r$ angles in Fig.~\ref{fig:vr-e}. We simulate the binaries as described in Sec.~\ref{sec:projection-effect}, and their eccentricities are drawn from a uniform distribution between 0 and 1.  We use an arbitrary binary separation because it does not play any role in Eq.~\ref{eq:vr-angle} and in the result. With all randomly sampled eccentricities, orbital phases, and orientations, we compute $\gamma$ using Eq.~\ref{eq:vr-angle-xy} in units of degree.

The left panel of Fig.~\ref{fig:vr-e} shows that the observed $v$-$r$ angle is strongly correlated with the eccentricity despite the random orbital phases and orientations. The distribution is symmetric with respect to $\gamma=90^\circ$\ due to the symmetry of the Keplerian orbit. The strong connection between eccentricity and the observed $v$-$r$ angle implies that the observed $v$-$r$ angle contains much of the information about the individual eccentricity even when the orbital phase and the binary orientation are not known.

Given the strong relation between eccentricity and the observed $v$-$r$ angle, it is not surprising that the observed $v$-$r$ angle distribution is directly tied to the underlying eccentricity distribution. The right panel of Fig.~\ref{fig:vr-e} gives a few examples of the observed $v$-$r$ angle distributions for different eccentricity distributions. For the cases where all wide binaries have the same eccentricity, the observed $v$-$r$ angle distribution is equivalent to the one-dimensional vertical slice of the left panel. For example, when all wide binaries have circular orbits ($e=0$), the observed $v$-$r$ angle distribution is strongly peaked at 90$^\circ$\ with extended tails due to the projection effects.

For a power-law eccentricity distribution $f_e(e)\propto e^{\alpha}$, the dependence of the observed $v$-$r$ angles on the exponent $\alpha$ is shown in Fig.~\ref{fig:vr-e}, right. A thermal eccentricity distribution ($\alpha=1$) leads to a uniform observed $v$-$r$ angle distribution. This interesting relation arises from the property of the thermal eccentricity distribution. The thermal eccentricity
distribution corresponds to a phase-space distribution function that only depends on orbital energy and not on angular momenta \citep{Jeans1919, Ambartsumian1937}. Therefore, its phase-space distribution function is isotropic, resulting in the flat observed $v$-$r$ angle distribution (Section \ref{sec:discussion}; S. Tremaine, priv. comm.). When $\alpha>1$, the so-called superthermal distribution, the observed $v$-$r$ angle distribution peaks at 0$^\circ$\ and 180$^\circ$, with a minimum at 90$^\circ$. Fig.~\ref{fig:vr-e} (right) shows an extreme case where $\alpha=4$. 

The right panel of Fig.~\ref{fig:vr-e} shows that different eccentricity distributions have distinctive observed $v$-$r$ angle distributions. In the rest of the paper, we develop a rigorous method to infer the eccentricity distribution from the observed $v$-$r$ angle distribution.

\subsection{Deriving the likelihood $p(\gamma|e)$}
\label{sec:p-gamma-bar-e}

Here we derive the likelihood of the $v$-$r$ angle given the eccentricity, $p(\gamma|e)$, which is critical for the Bayesian inference of the posterior $p(e|\gamma)$. We first consider a simplified case when the binary is viewed face-on (no projection effects), $\gamma_{3D}$ (Eq.~\ref{eq:vr-angle-0}). Let $\Gamma_{3D}=\cos \gamma_{3D}$, and 
\begin{equation}
    p(\Gamma_{3D}|e)= p(f|e) \left|\frac{df}{d\Gamma_{3D}}\right|.
\end{equation}
Because the mean anomaly is linear in time, its likelihood is uniform, $p(M|e) = (2\pi)^{-1}$. Therefore, we have 
\begin{equation}
\label{eq:p-f}
    p(f|e) = p(M|e) \left|\frac{dM}{df}\right|= \frac{1}{2\pi} \frac{(1 - e^2)^{3/2}}{(1+e \cos f)^2},
\end{equation}
where we use the chain rule and Eq.~\ref{eq:cosu}, \ref{eq:sinu}, \ref{eq:mean-anomaly} to write $dM/df$ as a function of $f$ in the last equality.

By differentiating Eq.~\ref{eq:vr-angle} with respect to $f$, we have
\begin{equation}
\label{eq:dfdgamma}
    \frac{df}{d\Gamma_{3D}} = \frac{(1 + e^2 + 2e\cos f)^{3/2}}{e (\cos f (1+e^2) + e (1+ \cos ^2 f))}.
\end{equation}

The relation between $\Gamma_{3D}$ and $\cos f$ is not single-valued. For example, for any eccentricity, both apogee and perigee have $\Gamma_{3D}=0$ but their $\cos f$ are different (1 and $-1$ respectively). Using Eq.~\ref{eq:vr-angle}, we can solve for the relation between $\cos f$ and $\Gamma_{3D}$ as
\begin{equation}
\label{eq:f-of-Gamma}
    \cos f = \frac{-\Gamma_{3D}^2 \pm \sqrt{\Gamma_{3D}^4-\Gamma_{3D}^2(1+e^2) + e^2}}{e},
\end{equation}
and we use $f_0$ and $f_1$ to represent the solutions with the plus and minus signs, correspondingly. Then the likelihood of $\Gamma_{3D}$ given an eccentricity is
\begin{equation}
\label{eq:p-Gamma-analytic}
\begin{split}
    & p(\Gamma_{3D}|e) = \\
    & p(f_0|e) \left| \left(\frac{df}{d\Gamma_{3D}} \right)_{f_0}\right| + p(f_1|e) \left| \left(\frac{df}{d\Gamma_{3D}} \right)_{f_1}\right|,
\end{split}
\end{equation}
where $p(f_0|e)$ and $p(f_1|e)$ are evaluated using Eq.~\ref{eq:p-f} and \ref{eq:f-of-Gamma}, and $df/d\Gamma_{3D}$ is evaluated using Eq.~\ref{eq:dfdgamma} and \ref{eq:f-of-Gamma}. With Eq.~\ref{eq:p-Gamma-analytic}, one has 
\begin{equation}
\label{eq:p-gamma-analytic}
p(\gamma_{3D}|e) = p(\Gamma_{3D}|e) \left| \sin \gamma_{3D} \right|.
\end{equation}

\begin{figure*}
	\centering
	\includegraphics[width=0.45\linewidth]{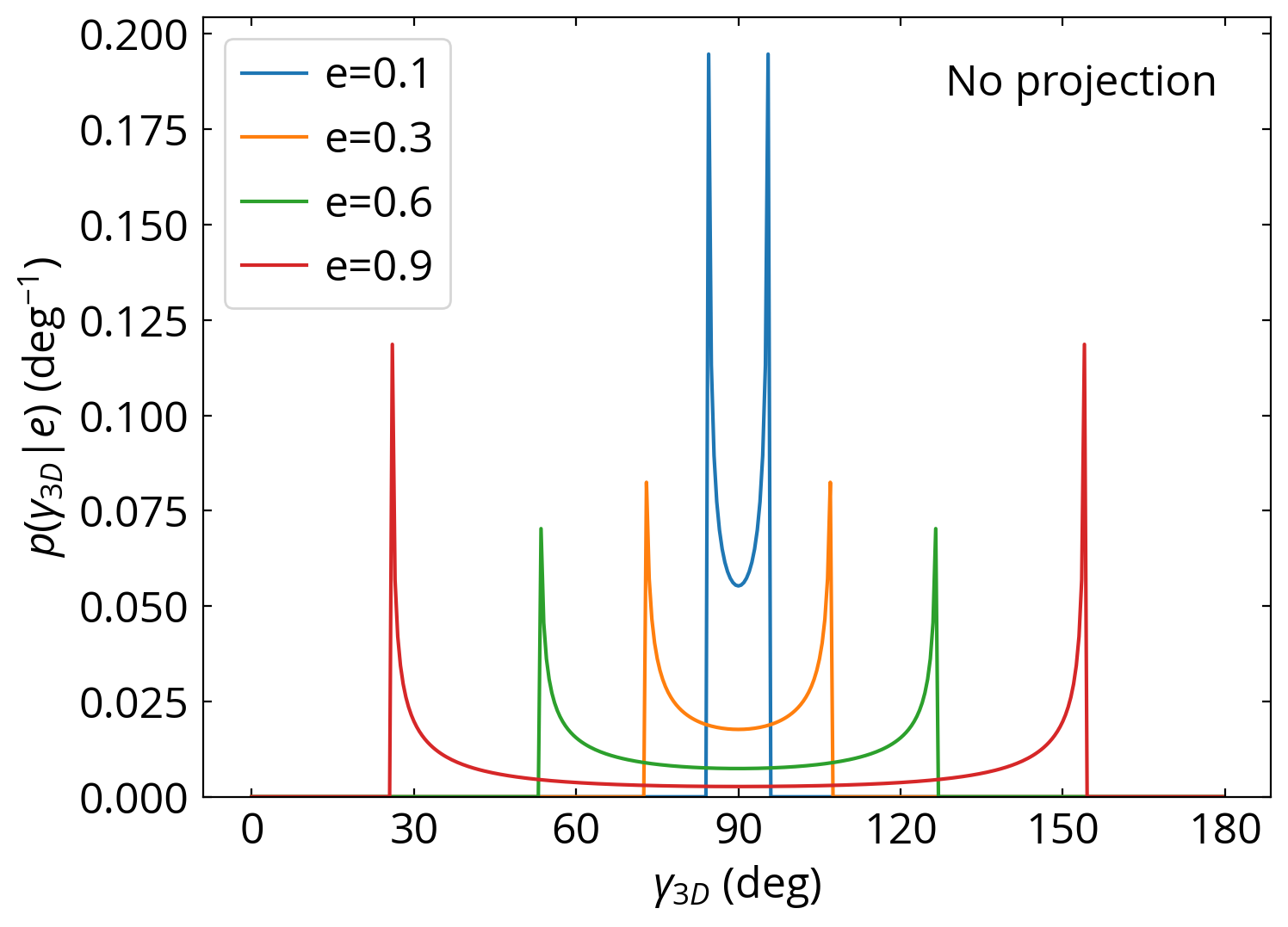}
	\includegraphics[width=0.45\linewidth]{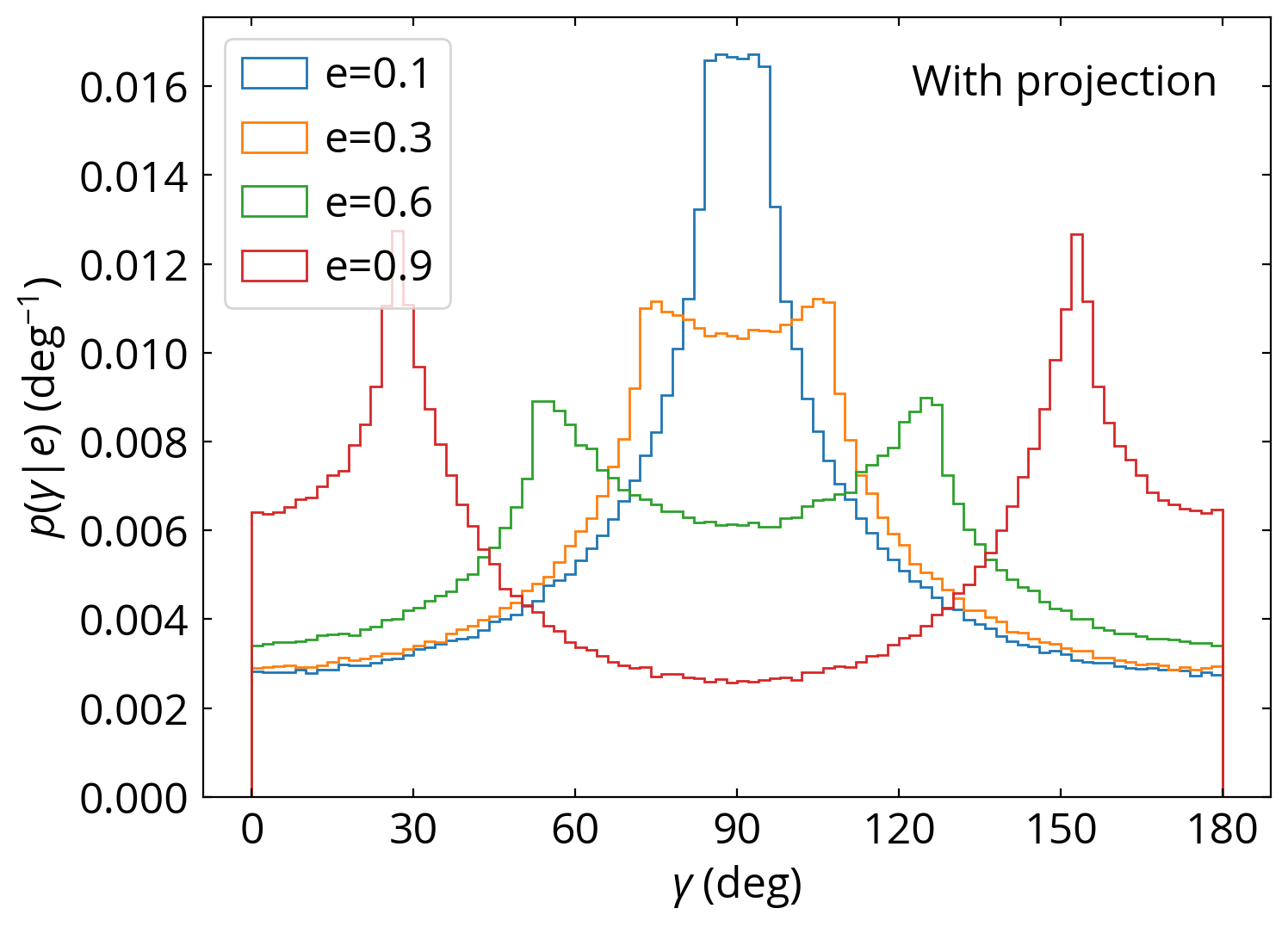}
	\caption{The distributions of $v$-$r$ angles for different eccentricities. Left: The distributions for $p(\gamma_{3D}|e)$, i.e. without on-sky projection effects. The lines are computed analytically using Eq.~\ref{eq:p-gamma-analytic}. Right: The simulated distributions for $p(\gamma|e)$, including the projection effects.}
	\label{fig:gamma-bar-e}
\end{figure*}

The resulting likelihoods $p(\gamma_{3D}|e)$ for different eccentricities are shown in the left panel of Fig.~\ref{fig:gamma-bar-e}. The overall behavior of $p(\gamma_{3D}|e)$ is, when $e=0$, $p(\gamma_{3D}|e=0)$ is a delta function at $\gamma_{3D}=90^\circ$. As $e$ becomes larger, the allowed range of $\gamma_{3D}$ becomes wider, and $p(\gamma_{3D}|e)$ peaks at the boundaries of $\gamma_{3D}$ (Fig.~\ref{fig:gamma-bar-e} left panel). From Eq.~\ref{eq:dfdgamma}, the allowed range of $\gamma_{3D}$ for a given $e$ is

\begin{equation}
\label{eq:cos-gamma-e}
    \left| \cos \gamma_{3D} \right| \le e.
\end{equation}
As we will see, this concise relation is still useful for observed $v$-$r$ angles ($\gamma$) even when the projection effects are present.

Including the projection effects makes an analytic expression of $p( \gamma|e)$ challenging. We define $\Gamma \equiv \cos \gamma$, and 
\begin{equation}
    p(\Gamma|e)= p(f|e) \left|\frac{df}{d\Gamma} \right|,
\end{equation}
where $df/d\Gamma$ can be computed by differentiating Eq.~\ref{eq:gamma-XY-ana} with respect to $f$. However, unlike Eq.~\ref{eq:f-of-Gamma} where $\cos f$ can be written as a function of $\Gamma_{3D}$, there seems no simple way to express $\cos f$ as a function of $\Gamma$ using Eq.~\ref{eq:gamma-XY-ana}. In Section \ref{sec:appendix}, we present formal expressions for $p(\Gamma|e)$ which can be evaluated numerically and compared to simulations in Fig. \ref{fig:vr-e}, right.

Alternatively, we obtain $p(\gamma|e)$ using numerical simulations. For every $e$ from 0 to 1 with a step of 0.01, we run a large number of simulated binaries with random orientation and orbital phase, and compute their observed $v$-$r$ angles using Eq.~\ref{eq:vr-angle-xy}. The right panel of Fig.~\ref{fig:gamma-bar-e} shows $p(\gamma|e)$ for selected eccentricities.

In contrast to $p(\gamma_{3D}|e)$ in the left panel of Fig.~\ref{fig:gamma-bar-e}, $p(\gamma|e)$ in the right panel spans all possible $\gamma$ from 0 to 180$^\circ$\ and does not have cutoffs at any certain $\gamma$ for any eccentricities due to the projection effects. For example, for a circular orbit $p(\gamma|e=0)$ peaks at $90^\circ$\ with long tails towards small and large $\gamma$. Interestingly, we find that the peaks of $p(\gamma|e)$ coincide perfectly with Eq.~\ref{eq:cos-gamma-e}. This result is not surprising because $p(\gamma_{3D}|e)$ strongly peaks at the boundaries, and $p(\gamma|e)$ is a convolution of $p(\gamma_{3D}|e)$ with projection effects and therefore peaks are preserved. We plot the relation $\left| \cos \gamma \right|=e$ as a solid black line in Fig.~\ref{fig:vr-e}, in excellent agreement with the peak. 

The fact that $p(\gamma|e)$ peaks at $\left| \cos \gamma \right|=e$ is particularly useful. It means that for a wide binary with an measured $\gamma$, we can compute its most probable eccentricity, i.e. the maximum of $p(e|\gamma)$, by using $e =\arccos \gamma$ under a uniform prior for $p(e)$. We present a more complete Bayesian procedure to obtain full $p(e|\gamma)$ in Sec.~\ref{sec:e-individual}.

\section{Measuring eccentricity for {\it Gaia} wide binaries}
\label{sec:observation}

\subsection{Notations for observable quantities}

We consider two stars with right ascension $\alpha_i$ and declination $\delta_i$, where the subscript $i=0, 1$ indicates the star. When two stars are close on the sky, we can find a two-dimensional Cartesian tangent plane defined by the direction of right ascension and declination direction. On this plane, the separation vector is 
\begin{equation}
\label{eq:s-vector}
\vec{s} = ((\alpha_1 - \alpha_0)\cos \delta, (\delta_1 - \delta_0)),
\end{equation}
where $\delta=(\delta_1 + \delta_0)/2$. Vector $\vec{s}$ is the separation between two stars' coordinates projected on the tangential Cartesian coordinate system, and it has units of angle.

The proper motion on the plane is $\vec{\mu}_i = (\mu_{\alpha^*, i}, \mu_{\delta, i})$, where $\mu_{\alpha^*, i} = \mu_{\alpha,i} \cos \delta_i$. The proper motion difference vector is defined as 
\begin{equation}
\label{eq:mu-vector}
\Delta \vec{\mu} = \vec{\mu}_1 - \vec{\mu}_0.
\end{equation}

The space velocity vector is related to the proper motion by $\vec{v}_i \propto \vec{\mu}_i / \texttt{parallax}$. Since the velocity vector has the same direction as the proper motion, we can measure the projected $v$-$r$ angle using the proper motions. In this case, the parallax measurements and their uncertainties do not play any role in the $v$-$r$ angle measurements.

The observed $v$-$r$ angle is computed by
\begin{equation}
\label{eq:vrangle-obs}
\gamma = \arccos \frac{\Delta \vec{\mu} \cdot \vec{s}}{|\Delta \mu| |s|}.
\end{equation}
We intentionally use the same notation $\gamma$ as in Eq.~\ref{eq:vr-angle-xy} because proper motions and separation vectors are two-dimensional vectors, and thus $\gamma$ is a projected quantity.

From Eq.~\ref{eq:vrangle-obs}, the uncertainty of $\gamma$ depends on the uncertainties of $\vec{s}$ and $\Delta \vec{\mu}$. Since the uncertainties in the separation vector $\vec{s}$ from {\it Gaia} are negligible (better than $0.1$\%), the uncertainty on $\gamma$ mainly comes from the uncertainty in the proper motion difference $\Delta \vec{\mu}$, which is (e.g. \citealt{El-Badry2021})
\begin{equation}
\sigma_{\Delta \mu} = \frac{1}{\Delta \mu}
 [(\sigma^2_{\mu^*_{\alpha,0}} + \sigma^2_{\mu^*_{\alpha,1}})\Delta\mu^2_\alpha +
 (\sigma^2_{\mu^*_{\delta,0}} + \sigma^2_{\mu^*_{\delta,1}})\Delta\mu^2_\delta
 ]^{1/2},
\end{equation}
where $\sigma_{\mu^*_{\alpha,0}}$ is the error of $\mu^*_{\alpha,0}$, $\sigma_{\mu_{\delta,0}}$ is the error of $\mu_{\delta,0}$, $\Delta\mu^2_\alpha=(\mu^*_{\alpha,1}-\mu^*_{\alpha,0})^2$, and $\Delta\mu^2_\delta=(\mu_{\delta,1}-\mu_{\delta,0})^2$. 

Error propagation of Eq.~\ref{eq:vrangle-obs} yields
\begin{equation}
\label{eq:gamma-error}
\sigma_\gamma \approx \frac{180}{\pi} \frac{\sigma_{\Delta \mu}}{\Delta \mu},
\end{equation}
where $\sigma_\gamma$ is the uncertainty of $\gamma$ in units of degrees. We use the symbol $\approx$ to indicate that this relation has a few assumptions. First, this relation assumes that $\Delta\mu_\alpha$ and $\Delta\mu_\delta$ are equal and mutually independent. Second, the error propagation assumes that the error distribution is Gaussian, but in reality the error distribution of $\gamma$ is truncated at 0$^\circ$\ and 180$^\circ$\ and hence is not Gaussian. Therefore, for $\gamma$ close to the boundaries, Eq.~\ref{eq:gamma-error} may overestimate the error by a factor up to 1.6. Also for $\sigma_{\Delta\mu}/\Delta\mu>1$ when the $v$-$r$ angle is poorly constrained, Eq.~\ref{eq:gamma-error} overestimates $\sigma_\gamma$ because the real error distribution is truncated. In general, these issues only mildly affect the uncertainty estimates and do not affect the main results. Therefore, for simplicity, we use Eq.~\ref{eq:gamma-error} to compute the uncertainties of $\gamma$.

Since wide binaries are identified as two co-moving stars, usually their $\Delta \vec{\mu}$ are small. If the proper motion difference is consistent with zero, then $\gamma$ would be poorly constrained. Furthermore, when ${\Delta \mu}/\sigma_{\Delta \mu}\gg1$, $\sigma_\gamma$ becomes $180/\sqrt{12}=52^\circ$, the standard deviation of a uniform distribution, and Eq.~\ref{eq:gamma-error} does not hold. For these reasons, we only consider wide binaries with proper motion differences inconsistent with zero at more than 3-sigma (${\Delta \mu}/\sigma_{\Delta \mu}>3$), where their $\gamma$ can be determined to a precision better than $20^\circ$\ (Eq.~\ref{eq:gamma-error}).

Given a wide binary $i$, we approximate the uncertainty distribution of $\gamma_i$ as a Gaussian distribution truncated at 0$^\circ$ and 180$^\circ$,
\begin{equation}
\label{eq:p-gamma-gamma}
p(\gamma_{true,i} | \gamma_{obs,i})=\frac{1}{Z} \exp \left( -\frac{(\gamma_{obs,i} - \gamma_{true,i})^2}{2 \sigma_{\gamma,i}^2} \right),
\end{equation}
where $\sigma_{\gamma,i}$ is evaluated from Eq.~\ref{eq:gamma-error}, $\gamma_{obs,i}$ is the measured $v$-$r$ angle, and $\gamma_{true,i}$ is the ground-truth $v$-$r$ angle. Because $\gamma$ is defined in a range between 0$^\circ$\ and 180$^\circ$, $p(\gamma_{true,i} | \gamma_{obs,i})=0$ for $\gamma_{true,i}<0^\circ$ or $\gamma_{true,i}>180^\circ$. The normalizing factor $Z$ ensures that $p(\gamma_{true,i} | \gamma_{obs,i})$ is normalized to 1, and $Z\ne \sigma_{\gamma,i} \sqrt{2\pi}$ because the Gaussian distribution here is truncated at 0$^\circ$ and 180$^\circ$.

\subsection{{\it Gaia} systematics in close pairs}
\label{sec:gaia-systematics}

We investigate if {\it Gaia} has any systematics in the $v$-$r$ angle measurements. We query {\it Gaia} data in a crowded region at Galactic latitudes between $5^\circ$ and $7^\circ$ and Galactic longitudes between $63^\circ$ and $65^\circ$. We require that all stars have parallaxes $>0$\,mas to avoid spurious astrometric solutions. Then we collect all pairs with angular separations $<10$\,arcsec and compute their observed $v$-$r$ angles using Eq.~\ref{eq:vrangle-obs}. To ensure robust $v$-$r$ angle measurements, pairs that have ${\Delta \mu}/\sigma_{\Delta \mu}<3$ are excluded. In this crowded field, most of the pairs are random pairs instead of wide binaries. We do not find significant differences in the $v$-$r$ angle distributions if we specifically remove wide binaries from the sample by requiring proper motion difference $>2$\masyr, or if we require astrometric quality indicators  {\texttt ruwe} $<1.4$.

\begin{figure}
	\centering
	\includegraphics[width=\linewidth]{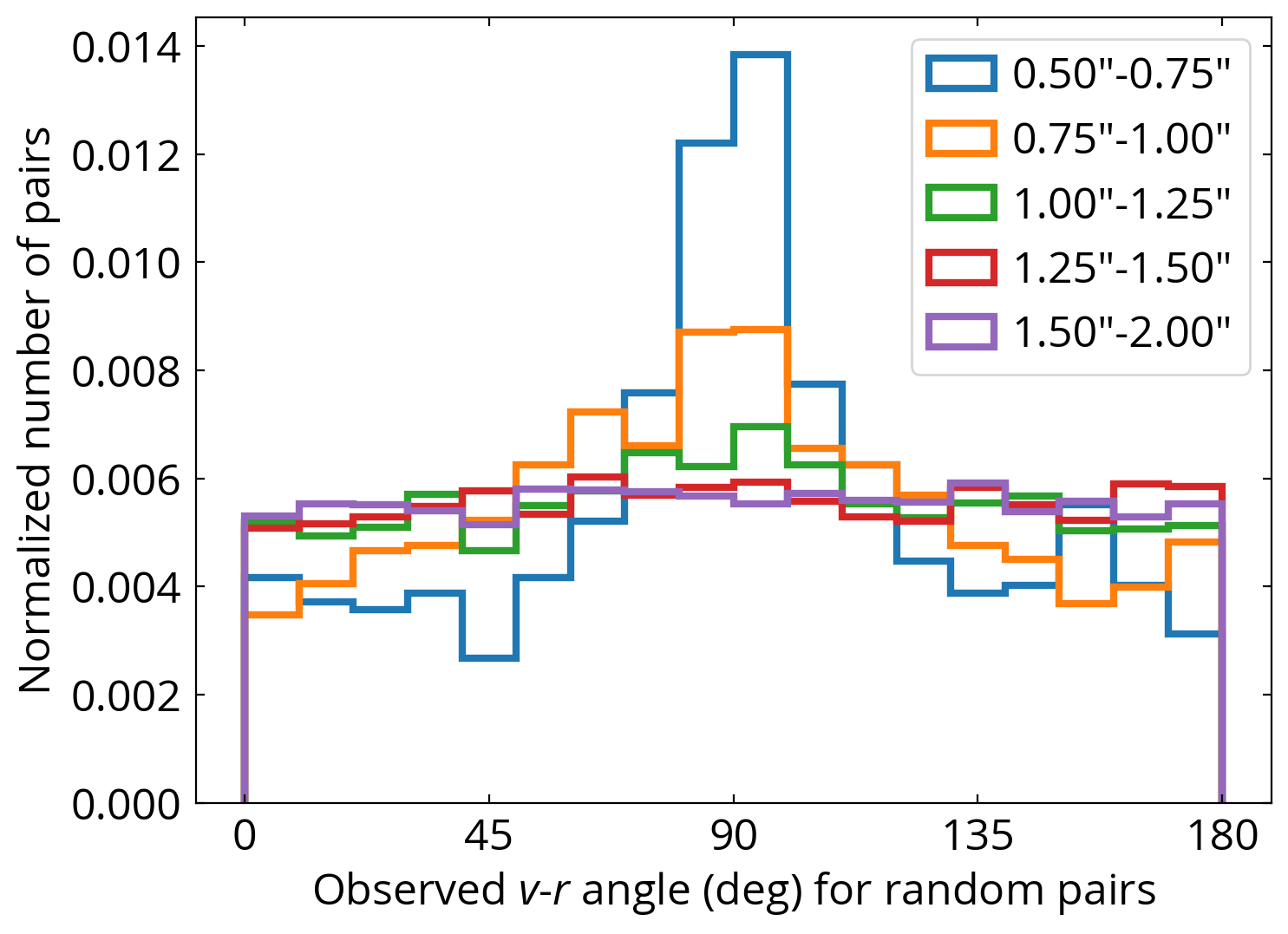}
	\caption{$v$-$r$ angle distributions for random pairs. The expected distribution is flat for random pairs. The enhanced peak at about 90$^\circ$ suggests that some {\it Gaia} systematics is present at pair separations below 1.25\,arcsec. }
	\label{fig:gaia-systematic}
\end{figure}

Fig.~\ref{fig:gaia-systematic} shows the observed $v$-$r$ angle distributions for random pairs. At separations $<1$\,arcsec, there is a prominent peak at  80$^\circ$ to 100$^\circ$. The peak is still present although weaker at 1-1.25\,arcsec. At separations $>1.25$\,arcsec, the $v$-$r$ angle distribution becomes flat, the expected distribution for random pairs. Therefore, at $<1.25$\,arcsec, there seems to be some (not easily identifiable) {\it Gaia} systematics that makes observed $v$-$r$ angles clustered around $90^\circ$.

In addition to the $v$-$r$ angle distributions, we find that the directions of the separation vectors (Eq.~\ref{eq:s-vector}) for these random pairs are not uniform at angular separations out to 2\,arcsec. This is likely due to the scanning law of {\it Gaia} such that pairs with separation vectors perpendicular to the scanning direction are more likely resolve. We also find a similar result in {\it Gaia} wide binaries from \cite{El-Badry2021} where the direction of separation vectors is only uniform at $>2$\,arcsec but not at $<2$\,arcsec. In this paper, we focus on $v$-$r$ angles and not the separation directions, and since it is unlikely that the binary orientation at separations of $\sim100$\,AU can be correlated with their eccentricities \citep{Hamilton2019,Hamilton2022}, thus the non-uniform separation direction at $<2$\,arcsec plays a minor role in our result.

These {\it Gaia} systematics mainly affects binaries with smallest separations, $\lesssim 100$\,AU in our sample. To ensure that our results are not affected by {\it Gaia}'s systematics on $v$-$r$ angles, we focus on wide binaries with separations $>1.5$\,arcsec in the following analysis.  

\subsection{$v$-$r$ angle measurements for {\it Gaia} wide binaries}
\label{sec:gaia-vr}

\begin{figure}
	\centering
	\includegraphics[width=\linewidth]{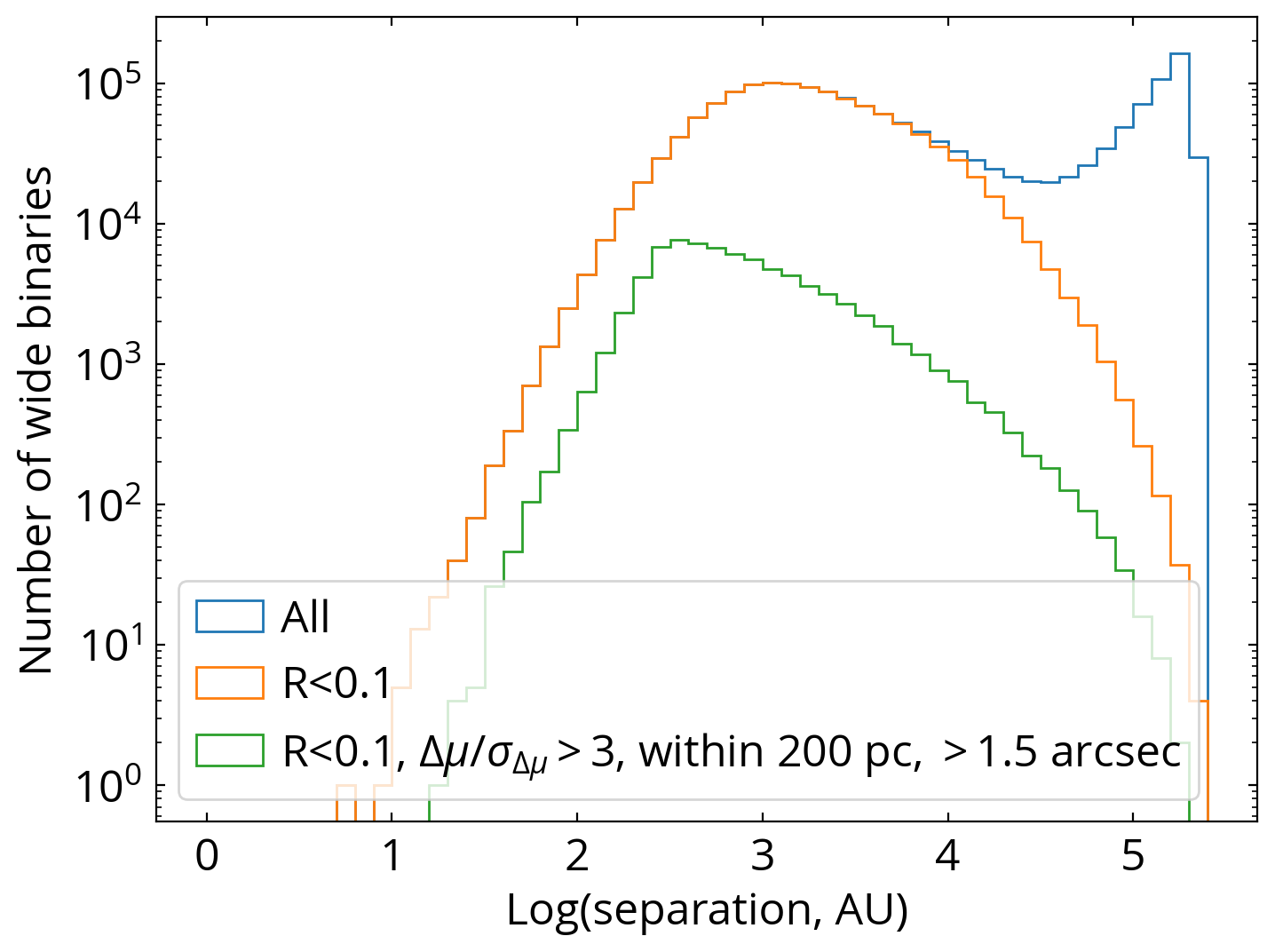}
	\caption{The separation distributions of wide binaries identified from {\it Gaia} EDR3 \citep{El-Badry2021}. The blue histogram has an enhanced number of wide binaries at large separation, and most of them are chance-alignment pairs. The orange histogram shows the numbers after we require the probability of being a chance alignment $R<0.1$. The green histogram is the number of wide binaries that have distances $<200$\,pc, angular separations $>1.5$\,arcsec, and non-zero proper motion differences detected at more than 3-$\sigma$.  }
	\label{fig:gaia-sep}
\end{figure}

We use Eq.~\ref{eq:vrangle-obs} to measure the $v$-$r$ angles for $\sim$1 million wide binaries within 1\,kpc identified from {\it Gaia} EDR3 \citep{El-Badry2021}. Resolved triples are not included in this catalog. The separation distribution of wide binary candidates is shown as the blue histogram in Fig.~\ref{fig:gaia-sep}. The enhanced number of wide binaries at separations $>10^5$\,AU is due to the chance-alignment pairs. To avoid these chance-alignment pairs, we use the parameter $\mathcal{R}$, the probability of a wide binary being a chance-alignment pair computed from \cite{El-Badry2021}. $\mathcal{R}$ is estimated by comparing the number of wide binaries with the number of chance-alignment pairs in the parameter space of the target wide binary, where the chance-alignment sample is constructed by doing the wide binary search after shifting stars' positions. Using $\mathcal{R}<0.1$ strongly reduces the chance alignment pairs at large separations (orange histogram). Furthermore, we require that the angular separations of wide binaries are $>1.5$\,arcsec to avoid {\it Gaia} systematics (Sec.~\ref{sec:gaia-systematics}).

The proper motion difference in {\it Gaia} EDR3 can be measured to a precision of $\sigma_{\Delta \mu}=0.1$\,\masyr, the median value in the catalog. This corresponds to a physical relative velocity of 0.5\kms\ (0.05\kms) at 1 (0.1) kpc, corresponding to the orbital velocity of a circular equal-solar-mass binary at a semi-major axis of $1\times10^4$ ($1\times10^6$) AU.

Since we apply the signal-to-noise ratio (SNR) cut on the proper motion difference (${\Delta \mu}/\sigma_{\Delta \mu}>3$) to ensure reliable $v$-$r$ angle measurements, this criterion preferentially removes eccentric orbits because eccentric orbits stay longer at larger separations with the lower orbital velocity compared to the less eccentric orbits. For wide binaries at 200\,pc, the median $\sigma_{\Delta \mu}$ in the sample is $0.1$\,\masyr, and a 3-$\sigma$ cut of $0.3$\,\masyr corresponds to a physical velocity of 0.28\kms. Assuming random binary orientations and an eccentricity distribution $f_e(e)\propto e^{0.5}$, we find that this physical velocity precision is able to recover the orbital motions of 91\% (44\%) binaries with separations at $10^3$ ($10^4$)\,AU.  Therefore, when analyzing the eccentricity distribution of wide binaries, we limit the sample to distances within 200\,pc (parallaxes of the primary $>5$\,mas) to reduce potential bias. In Sec.~\ref{sec:systematics}, we investigate this selection effect in more detail.

Limiting the sample to 200\,pc also reduces the mass dependence in our results. At 200\,pc, {\it Gaia}'s magnitude limit of $20$\,mag can detect sources with absolute $G$-band magnitude down to 13.5\,mag, which includes most of the main-sequence stars except for the fainest M-dwarfs. Faint old white dwarfs may not be detected at 200\,pc, but white-dwarf wide binaries only comprise $\sim1$\% of the wide binary sample \citep{El-Badry2021} and plays a minor role in our results. Therefore, although binaries with smaller separations are closer due to {\it Gaia}'s spatial resolution limit, the mass distribution is similar across the binary separations investigated here because {\it Gaia} detect most of the main-sequence stars within 200\,pc.

\begin{figure}
	\centering
	\includegraphics[width=\linewidth]{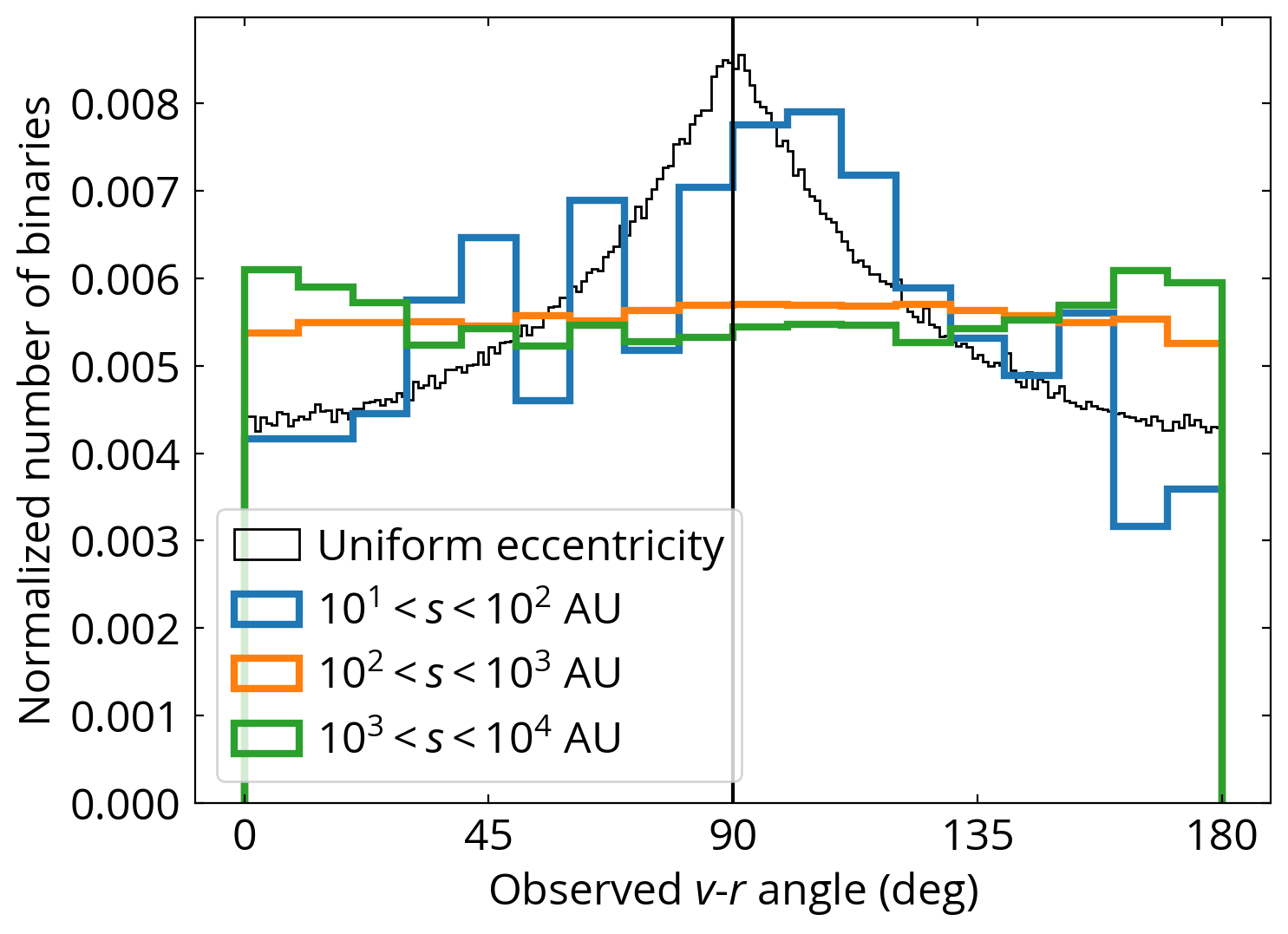}
	\caption{Distributions of observed $v$-$r$ angles for different binary separations. The background black histogram shows the simulated distribution for a uniform eccentricity distribution. The thermal eccentricity distribution has a flat $v$-$r$ angle distribution. The observed $v$-$r$ angle distributions qualitatively show that the eccentricity distribution of wide binaries at $\sim100$\,AU is close to uniform (blue). In contrast, wide binaries at $10^3$-$10^4$\,AU have enhanced numbers at at 0$^\circ$ and 180$^\circ$ (green), indicating a superthermal eccentricity distribution.}
	\label{fig:gaia-vrangle}
\end{figure}

Fig.~\ref{fig:gaia-vrangle} shows the distributions of the observed $v$-$r$ angles for {\it Gaia} EDR3 wide binaries within 200\,pc for three different ranges of binary separations projected on the sky, $s$. Binaries with $s<100$\,AU (blue) show an enhancement around 90$^\circ$, qualitatively similar to that predicted for an underlying uniform eccentricity distribution (black histogram). Wide binaries at separations between $10^2$ and $10^3$\,AU (orange) have a nearly flat distribution, suggesting a thermal eccentricity distribution (Fig.~\ref{fig:vr-e}). Interestingly, wide binaries at $10^3$-$10^4$\,AU (green) have enhanced numbers at 0$^\circ$ and 180$^\circ$, indicating the presence of highly eccentric orbits. Therefore, Fig.~\ref{fig:gaia-vrangle} shows that the eccentricity distribution evolves from uniform to thermal, and then super thermal with increasing binary separations.

\subsection{Bayesian inference for the eccentricity distribution}
\label{sec:e-dist}

Fig.~\ref{fig:gaia-vrangle} suggests that a single-parameter family of eccentricity distributions where the parameter depends primarily on the separation provides a good description of the data. Therefore, we adopt a functional form for the eccentricity distribution:
\begin{equation}
\label{eq:p-e-bar-alpha}
    p(e|\alpha) = (1+\alpha) e^{\alpha},
\end{equation}
where $\alpha$ is the parameter to be determined by the Bayesian inference. We only consider $\alpha>-1$ because $p(e|\alpha)$ cannot be normalized for $\alpha\le-1$. $p(e|\alpha)$ is a one-parameter family of functions that includes two important cases -- a uniform distribution (with $\alpha=0$) and a thermal eccentricity distribution (with $\alpha=1$). In Appendix~\ref{sec:generalized-e}, we discuss other functional form (e.g. multi-step function) for eccentricity distributions.

We use Bayesian inference to obtain the best fit for the parameter $\alpha$ given $v$-$r$ angle distribution $\{\gamma_{obs,i}\}$ for different separation bins. According to the Bayes' theorem, $p(\alpha | \{\gamma_{obs,i}\}) \propto p( \{\gamma_{obs,i}\}|\alpha)p(\alpha)$, and we adopt an uninformative prior for $p(\alpha)$. Because every wide binary is independent, $p(\{\gamma_{obs,i}\}|\alpha) = \prod_i p(\gamma_{obs,i}|\alpha)$. Then we can marginalize over $e$ by $p(\gamma_{obs,i}|\alpha)=\int p(\gamma_{obs,i}|e_i)p(e_i|\alpha) de_i$. After marginalizing over the uncertainties of $\gamma_{obs}$, we arrive the final Bayesian model as
\begin{equation}
\label{eq:baye-e-dist}
\begin{multlined}
p(\alpha | \{\gamma_{obs,i}\}) \propto \\ \Pi_i \int p(\gamma_{obs,i} | \gamma_{true,i}) p(\gamma_{true,i}| e_i)  p(e_i|\alpha) d\gamma_{true,i} de_i,
\end{multlined}
\end{equation}
where index $i$ refers to an individual wide binary. $p(\gamma_{obs,i} | \gamma_{true,i}) \propto p(\gamma_{true,i} | \gamma_{obs,i}) p(\gamma_{obs,i})$ and we use a flat prior for $p(\gamma_{obs,i})$ and $p(\gamma_{true,i} | \gamma_{obs,i})$ is from Eq.~\ref{eq:p-gamma-gamma} that incorporates the measurement uncertainties of $v$-$r$ angles in the model. $p(\gamma_{true,i}| e_i)$ is numerically derived in Sec.~\ref{sec:p-gamma-bar-e}, and $p(e_i|\alpha)$ is Eq.~\ref{eq:p-e-bar-alpha}. 

While the visual comparison between the observed distributions and the model in Fig.~\ref{fig:gaia-vrangle} is qualitatively useful, the observed distributions are affected the uncertainties of $v$-$r$ angles up to 20$^\circ$, which makes the observed distributions flatter than what we would have measured if $v$-$r$ angles were known to infinite precision. The Bayesian model in Eq.~\ref{eq:baye-e-dist} provides a tractable procedure that incorporates the measurement uncertainties to derive $\alpha$ for a given $v$-$r$ angle distribution. 

In principle, $\alpha$ can be a function of several physical parameters, including stellar ages and masses. In this work, we focus on its relation with binary separation. We bin {\it Gaia} wide binaries by logarithmic projected binary separations from $10^{1.5}$ to $10^{4.5}$\,AU with a bin size of 0.25 or 0.5\,dex, depending on the sample size. For each bin, we use Eq.~\ref{eq:baye-e-dist} to obtain $p(\alpha|\{\gamma_{obs,i}\})$ at different separation bins. We numerically compute the two-dimensional integral in Eq.~\ref{eq:baye-e-dist} using equal spacings of $\Delta \gamma_{true, i}=1$\,deg and $\Delta e_i=0.01$. The choice of spacings is a balance between the numerical accuracy and the efficiency because the two-dimensional integral needs to be computed for all wide binaries. $p(\alpha|{\gamma_{obs,i}})$ is evaluated for $\alpha$ from $-0.99$ to $3$ with a step of 0.01. With $p(\alpha|\{\gamma_{obs,i}\})$, we compute the most probable $\alpha_{best}$ where the maximum of $p(\alpha|\{\gamma_{obs,i}\})$ occurs, and compute the highest posterior density interval $[\alpha_0, \alpha_1]$ (i.e. the narrowest interval) that includes $68$ per cent of the area. We express the measurement and the uncertainty of $\alpha$ as $(\alpha_{best}) ^{\alpha_1-\alpha_{best}}_{\alpha_0 - \alpha_{best}}$. 

\begin{figure*}
	\centering
	\includegraphics[width=0.7\linewidth]{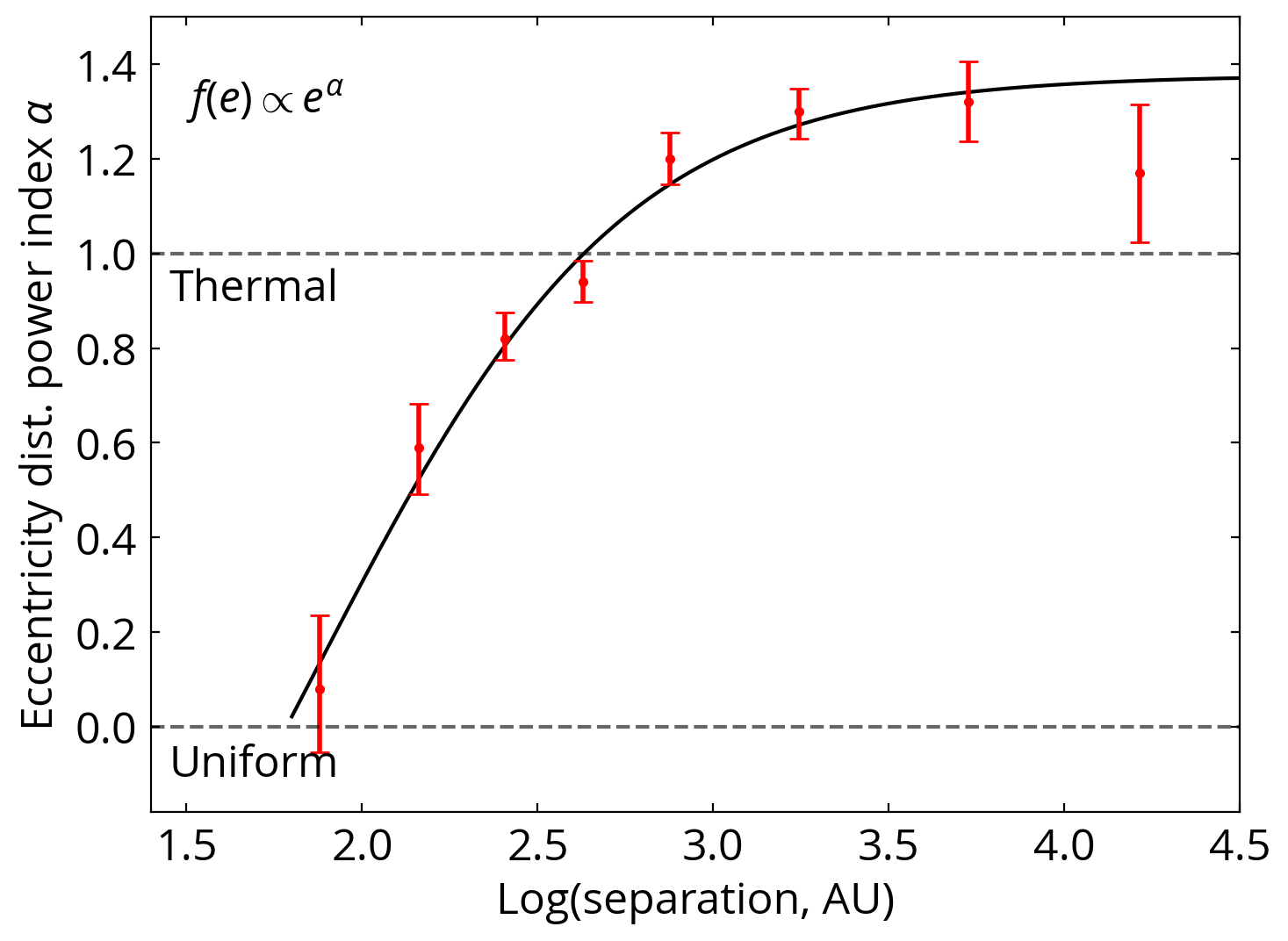}
	\caption{The power index of the eccentricity distribution $\alpha$ ($f_e(e)\propto e^{\alpha}$) as a function of binary separation. The horizontal dashed lines mark the uniform ($\alpha=0$) and thermal ($\alpha=1$) eccentricity distribution. The eccentricity distribution changes from a uniform distribution at $100$\,AU, to a thermal distribution at $\sim10^{2.7}$\,AU, and to a superthermal ($\alpha>1$) at $>10^{3}$\,AU. The black line shows the best-fit relation.  }
	\label{fig:alpha-sep}
\end{figure*}

\begin{figure}
	\centering
	\includegraphics[width=\linewidth]{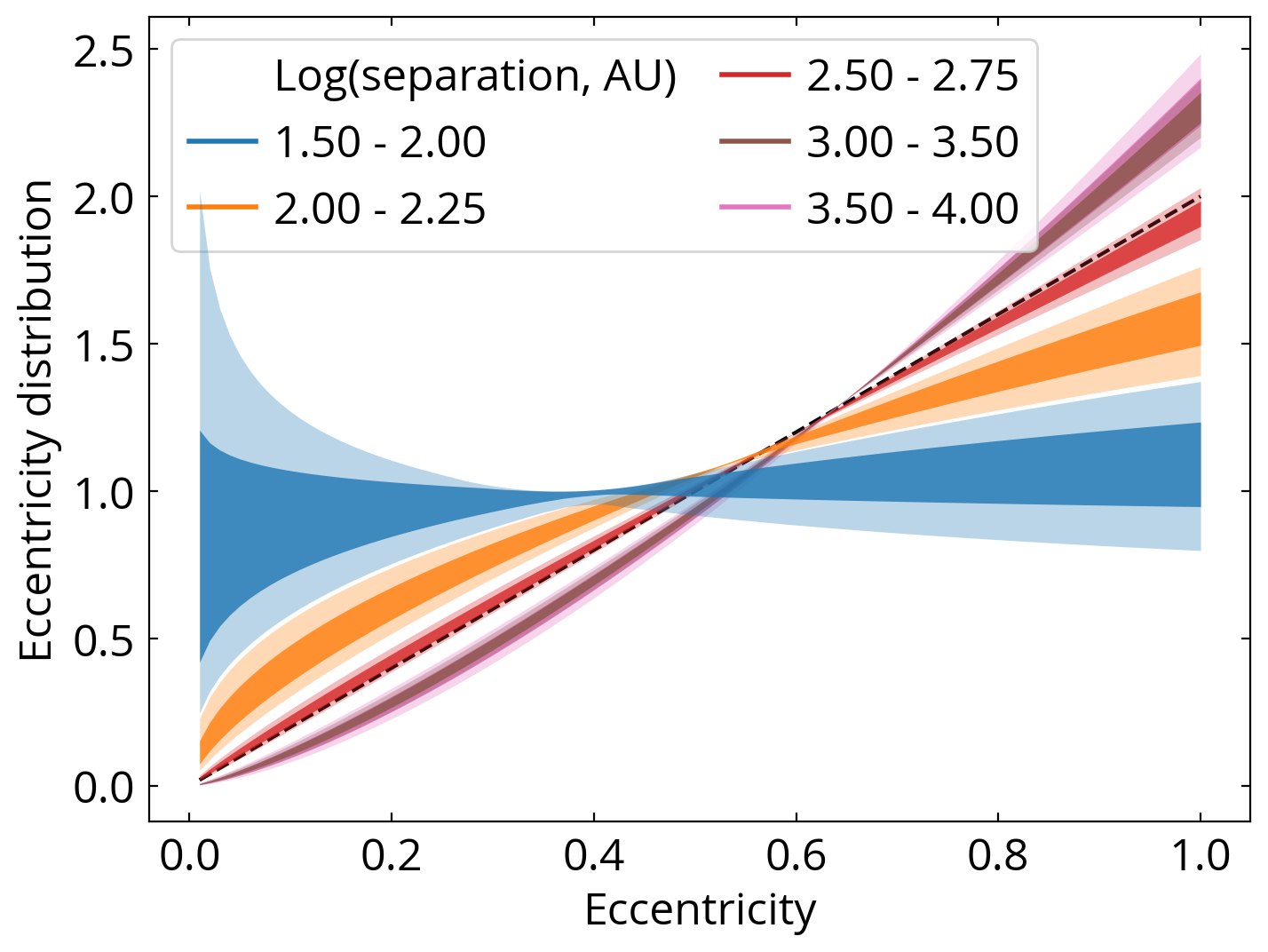}
	\caption{The eccentricity distributions for different binary separations. The dark and light shaded regions represent 1-$\sigma$ and 2-$\sigma$ uncertainties of the power-law index $\alpha$, respectively. The dashed black line shows the thermal eccentricity distribution ($f_e(e) =2e$).}
	\label{fig:e-dist}
\end{figure}

Fig.~\ref{fig:alpha-sep} shows the measured eccentricity distributions as a function of binary separations. The numerical values are tabulated in Table~\ref{tab:alpha-values}. At separations $<10^{2}$\,AU, $\alpha$ is consistent with zero, indicating a uniform eccentricity distribution. At separations $>10^{3}$\,AU, the eccentricity distribution becomes superthermal ($\alpha=1.32^{+0.09}_{-0.08}$), which explains the enhanced numbers close to 0$^\circ$ and 180$^\circ$\ in their $v$-$r$ angle distribution in Fig.~\ref{fig:gaia-vrangle}. Fig.~\ref{fig:e-dist} presents the realization of the eccentricity distributions at different binary separations.

\begin{table}[]
	\caption{Numerical values of $\alpha$ as a function of binary separations.}
	\label{tab:alpha-values}
	\center
	\begin{tabular}{ccc}
		\hline \hline
		Separation (log AU) & Number of binaries & $\alpha$ \\
		\hline
		$[1.50, 2.00]$ &686 & $0.08^{+0.15}_{-0.13}$ \\
        $[2.00, 2.25]$ &2814 & $0.59^{+0.09}_{-0.10}$ \\
        $[2.25, 2.50]$ &12357 & $0.82^{+0.05}_{-0.04}$ \\
        $[2.50, 2.75]$ &18379 & $0.94^{+0.04}_{-0.04}$ \\
        $[2.75, 3.00]$ &14884 & $1.20^{+0.05}_{-0.05}$ \\
        $[3.00, 3.50]$ &18496 & $1.30^{+0.05}_{-0.06}$ \\
        $[3.50, 4.00]$ &7570 & $1.32^{+0.09}_{-0.08}$ \\
        $[4.00, 4.50]$ &2301 & $1.17^{+0.14}_{-0.15}$ \\
		\hline \hline     
	\end{tabular}
\end{table}

\subsection{Bayesian inference for eccentricities of individual wide binaries}
\label{sec:e-individual}

Here, we use the eccentricity distributions derived from Sec.~\ref{sec:e-dist} as a prior to derive the eccentricity for individual wide binaries \citep{Tokovinin2020a}. The Bayesian inference for the eccentricity of an individual wide binary with an index $i$ is
\begin{equation}
\label{eq:baye}
p(e_i| \gamma_{obs,i}) \propto p(\gamma_{obs,i}| e_i) p(e_i),
\end{equation}
where $p(e_i|\gamma_{obs,i})$ is the likelihood of eccentricity $e_i$ given the observed $\gamma_{obs,i}$, $p(\gamma_{obs,i}|e_i)$ is the likelihood of $\gamma_{obs,i}$ given $e_i$, and $p(e_i)$ is the prior for the eccentricity distribution. Marginalizing over the measurement uncertainties of $\gamma_{obs,i}$, we have 
\begin{equation}
    \label{eq:baye-error}
     \begin{multlined}
p(e_i| \gamma_{obs,i}) \propto \\ \int p(\gamma_{true,i}| e_i) p(\gamma_{obs,i} | \gamma_{true,i}) d\gamma_{true,i}\ p(e_i),
\end{multlined}
\end{equation}
where $p(\gamma_{obs,i} | \gamma_{true,i})$ is the error distribution from Eq.~\ref{eq:p-gamma-gamma}.

As we show in Sec.~\ref{sec:e-dist}, the eccentricity distribution is a function of binary separation. Therefore, in contrast with Sec.~\ref{sec:e-dist} where we do not have prior knowledge about the population eccentricity distribution and use a flat prior for $p(\alpha)$, here we have an eccentricity prior $p(e_i)$ for an individual wide binary depending on its binary separation. This prior is $p(e_i)=(1+\alpha_{fit}) e_i^{\alpha_{fit}}$ and we adopt a functional form for $\alpha_{fit}$:
\begin{equation}
\label{eq:alpha-best-fit}
    \alpha_{fit}(w_i) = A \tanh ((w_i - B)/C) + D,
\end{equation}
where $w_i$ is the base-10 logarithm of projected binary separation (AU). This function asymptotes to $\alpha(w_i)=D+A$ at large separations ($w_i\gg B$) and to $\alpha(w_i)=D-A$ when $w_i\ll B$, with $B$ and $C$ parametrizing the location and the width of the transition region. The hyperbolic tangent fitting function is merely an empirical description of the overall trend and does not have physical motivations. The black line in Fig.~\ref{fig:alpha-sep} shows the best fit with $A=1.25$, $B=1.87$, $C=0.88$, and $D=0.12$. Due to the lack of constraints at small binary separations, we set $\alpha_{fit}=0$ at $w_i<1.78$ where Eq.~\ref{eq:alpha-best-fit} would give negative $\alpha_{fit}$.

For each {\it Gaia} wide binary, we use Eq.~\ref{eq:baye-error} to obtain the posterior of eccentricity $p(e_i|\gamma_{obs,i})$ using the separation-dependent eccentricity prior from Eq.~\ref{eq:alpha-best-fit}. Then we measure the most probable value and the highest posterior density interval that includes 68 per cent of the area. Table~\ref{tab:e-values} explains the entries of the electronic table that catalogs these measurements. For completeness, some entries are from \cite{El-Badry2021} with the same entry names. The entry \texttt{dpm\_sig} is computed using Eq.~\ref{eq:gamma-error} and is overestimated for $\sigma_{\Delta\mu}/\Delta\mu>1$ when $v$-$r$ angle is poorly constrained. In Sec.~\ref{sec:e-dist}, we limit our sample to $<200$\,pc so that the eccentricity distribution is not biased. When inferring the individual eccentricities here, we apply this approach for wide binaries at all distances. 

\begin{figure}
	\centering
	\includegraphics[width=\linewidth]{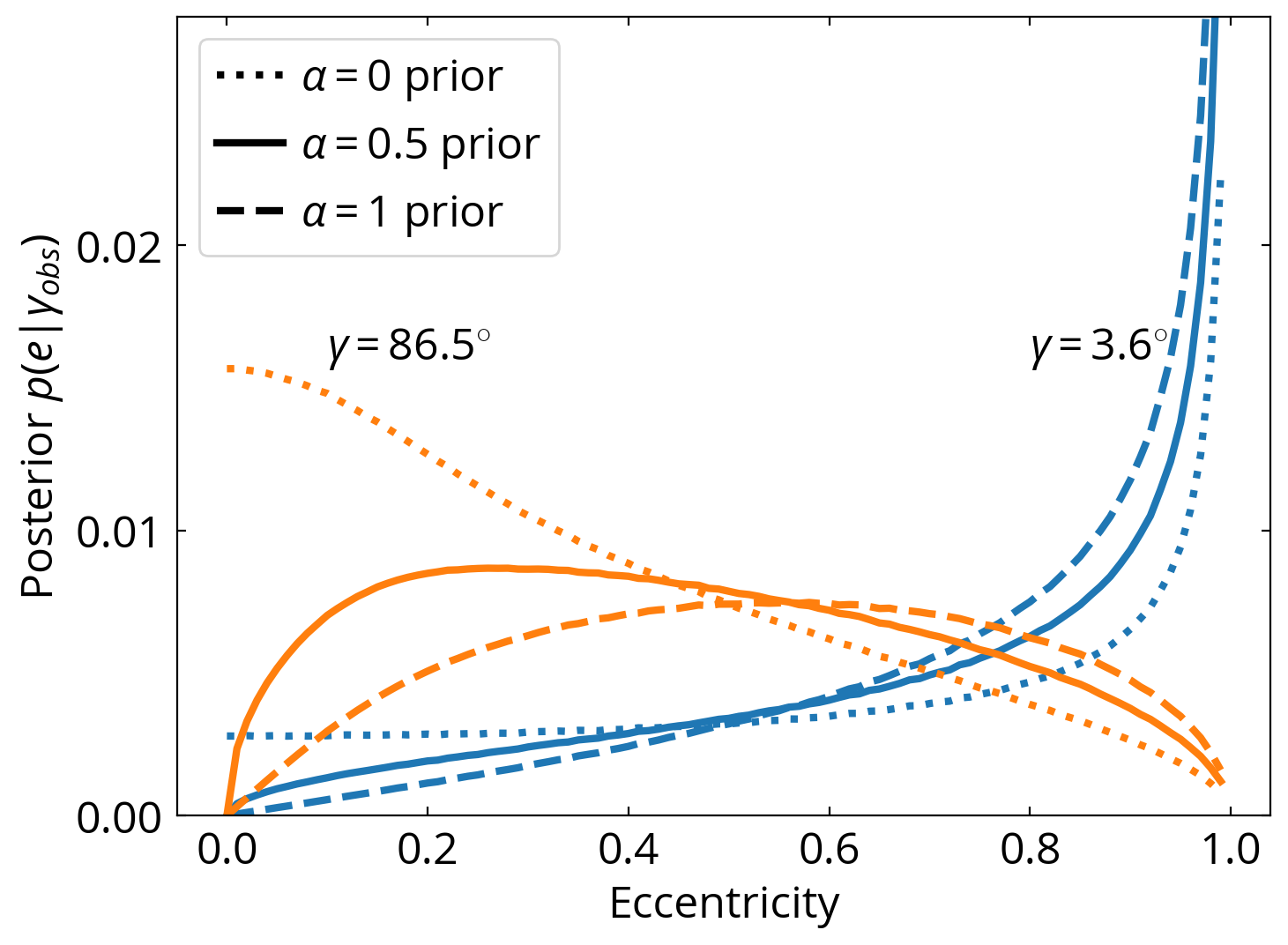}
	\caption{Examples of the eccentricity posterior $p(e|\gamma_{obs})$ for different $v$-$r$ angles ($\gamma$) and eccentricity distribution priors. For $\gamma$ close to 0$^\circ$ and 180$^\circ$ (blue), the eccentricity posterior peaks at $e=1$, regardless of the priors. The inferred eccentricity for $\gamma\sim90^\circ$\ strongly depends on the prior, with the most probable value shifting from 0 for the uniform eccentricity prior ($\alpha=0$, dotted orange line) to $0.58$ for the thermal eccentricity prior ($\alpha=1$, dashed orange line). } 
	\label{fig:e-pdf}
\end{figure}

Fig.~\ref{fig:e-pdf} shows how the posterior $p(e_i|\gamma_{obs,i})$ computed from Eq.~\ref{eq:baye-error} behaves for different $v$-$r$ angles and priors. We select two examples so that they both have $\alpha_{fit}\sim0.5$, with different $v$-$r$ angle measurements. To demonstrate how eccentricity priors $p(e_i)$ affect the posterior, we plot the eccentricity posteriors that use the uniform eccentricity prior ($\alpha=0$) and the thermal eccentricity prior ($\alpha=1$). For $\gamma$ close to 0$^\circ$ and 180$^\circ$, the eccentricity posterior strongly peaks at $e=1$. Different priors only affect the posterior tail toward $e=0$ without varying the most probable values much. For $\gamma$ close to 90$^\circ$, the eccentricity posterior is broad. The most probable eccentricities are strongly dependent on the prior, changing from 0 for the uniform prior to 0.58 for the thermal prior. 

The qualitative properties of eccentricity posteriors from Fig.~\ref{fig:e-pdf} are applicable to all other binary separations. In general, $\gamma$ close to 0$^\circ$ and 180$^\circ$\ has measured eccentricity close to $e=1$ and has smaller uncertainties, weakly dependent on the prior. For $\gamma$ close to 90$^\circ$, the inferred eccentricity strongly depends on the eccentricity prior and therefore on the binary separation. Therefore, it is important to include the separation dependence when inferring the eccentricity for individual wide binaries. The median 1-$\sigma$ uncertainties are 0.19 and 0.27 for binaries (with ${\Delta \mu}/\sigma_{\Delta \mu}>3$) with $v$-$r$ angles close to 0$^\circ$ and 90$^\circ$, respectively.

\begin{table*}[]
	\centering
	\caption{Descriptions for the catalog of individual wide binary eccentricities.}
	\label{tab:e-values}
	\begin{tabular}{ll} 
		\hline \hline
		Field 							& Description \\
		\hline
		\texttt{source\_id1}   		  & {\it Gaia} EDR3 source\_id of the primary \\
		\texttt{ra1} 						    & Right ascension of the primary from {\it Gaia} EDR3 (J2016.0; deg) \\
		\texttt{dec1} 						   & Declination of the H3 star from {\it Gaia} EDR3 (J2016.0; deg) \\
		\texttt{source\_id2}    		  & {\it Gaia} EDR3 source\_id of the secondary \\
		\texttt{ra2} 						    & Right ascension of the secondary from {\it Gaia} EDR3 (J2016.0; deg) \\
		\texttt{dec2} 						   & Declination of the secondary from {\it Gaia} EDR3 (J2016.0; deg) \\
		\texttt{sep\_AU}\tablenotemark{a} 						   & Projected binary separation (AU) \\
		\texttt{R\_chance\_align}\tablenotemark{a} 						   & Probability of being a chance-alignment pair \\
		\texttt{vr\_angle} 						   & Measured $v$-$r$ angle (deg) \\
		\texttt{vr\_angle\_error} 						   & Uncertainty of $v$-$r$ angle (deg) \\
		\texttt{dpm\_sig} 						   & The significance of proper motion difference ${\Delta \mu}/\sigma_{\Delta \mu}$ (unitless) \\
		\texttt{alpha} 						   & The power index used for the prior eccentricity distribution (unitless) \\
		\texttt{e} 						   & The most probable eccentricity (unitless) \\
		\texttt{e0} 						   & The lower eccentricity limit of the 68\% credible interval (unitless) \\
		\texttt{e1} 						   & The upper eccentricity limit of the 68\% credible interval (unitless) \\
		\hline \hline
		
	\end{tabular}
	\tablenotetext{a}{Entries from \cite{El-Badry2021} for completeness. }
\end{table*}

\subsection{Possible systematics}
\label{sec:systematics}

Since chance-alignment pairs have random $\vec{v}$ and $\vec{r}$ directions, the contamination in the wide binary sample from chance-alignment pairs would make the observed $v$-$r$ angle distribution flat, mimicking the thermal eccentricity distribution. Using the estimated contamination rate $\mathcal{R}$ for individual wide binaries, we find the contamination rates are $<3$\% for the samples at $s>10^4$\,AU, and are $<0.3$\% for the samples at $s<10^4$\,AU. Therefore, the contamination rate is low in our sample, and the flat $v$-$r$ angle distributions at large separations are physical and not due to the contamination from chance-alignment pairs.

The mean angular separation is 2.3\,arcsec for the $10^{1.5}$-$10^{2}$\,AU sample, 14.1\,arcsec for the $10^{3.0}$-$10^{3.5}$\,AU sample, and 135.1\,arcsec for the $10^{4.0}$-$10^{4.5}$\,AU sample. If the reported proper motion uncertainties are underestimated by the presence of nearby stars within 2\,arcsec, then the $10^{1.5}$-$10^{2}$\,AU sample would have more noisy $v$-$r$ angle measurements, resulting in a more random $v$-$r$ angle distribution and thus mimicking the thermal eccentricity distribution. In contrast, our result shows that the eccentricity distribution of $10^{1.5}$-$10^{2}$\,AU is close to uniform, in the opposite direction to this potential systematics. 

At large angular separations, the curvature of the sky becomes important and may affect the projection effect of wide binaries. This effect plays an important role when using wide binaries to test gravity theory in the low-acceleration regime \citep{Banik2018, Pittordis2019,El-Badry2019b}. We find that for binary separations and distances of our sample, this effect is $\ll 1$\,deg, well within our $v$-$r$ angle uncertainties. Therefore, the curvature-related projection effect plays a minor role in our results.  

We use the projected separations to bin the wide binaries, and projected separations are not equal to semi-major axis due to projection effects and time-averaging. For a randomly oriented three-dimensional vector, the projection effects cause the projected length on the $x$-$y$ plane to be reduced by a factor of $\pi/4=0.7854$. This projection effect does not explicitly depend on eccentricity.

The time-averaging effect is that, for a fixed semi-major axis, the separation (without projection) averaged over time is a function of eccentricity. Specifically, a more eccentric orbit stays longer at larger separations due to the smaller orbital velocity. From Eq.~\ref{eq:1}, the time-averaged separation of a face-on orbit is 
\begin{equation}
    \langle s \rangle = \frac{2+e^2}{2}a.
\end{equation}
For a circular orbit, $\langle s \rangle =a$ as expected. For an eccentric orbit with $e$ close to 1, its time-averaged separation is larger than the semi-major axis by a factor up to 1.5. Therefore, the time-averaging effect makes the observed separation larger than the semi-major axis for eccentric orbits.

Due to this time-averaging effect, when we use projected separations to bin the sample, we tend to select eccentric wide binaries with semi-major axes smaller than those of less eccentric binaries. Then because of the decreasing separation distribution at $>100$\,AU \citep{Raghavan2010}, there are more eccentric binaries scattering in a projected separation bin than scattering out, making the eccentricity distribution more eccentric than that of a sample binned by semi-major axes.

Another important selection effect is that we can only measure the $v$-$r$ angle when there is a significant non-zero proper motion difference. For wide binaries with the same semi-major axes and distances, this selection criterion preferentially excludes eccentric wide binaries because they stay a larger fraction of the orbit with lower orbital velocities, and hence smaller proper motion differences. Therefore, in contrast with the time-averaging effect that makes the eccentricity distribution more eccentric, this $v$-$r$ angle SNR criterion makes the distribution less eccentric.

While these selection effects affect the underlying eccentricity distribution of the wide binary sample, they affect the observed $v$-$r$ angles in a more complicated manner. The reason is that these selection effects also affect the binary orientation distribution and the orbital phase distribution. For example, the time-averaging effect would select eccentric binaries more at their apocenter and more at their face-on orientation. In contract, the $v$-$r$ angle SNR criterion more likely to select binaries at pericenter with face-on orientations due to their larger projected velocities. Therefore, although these effects change the underlying eccentricity distributions, they do not change the $v$-$r$ angle distribution like the relation between eccentricity and $v$-$r$ angle demonstrated in Sec.~\ref{sec:p-gamma-bar-e} where we consider random binary orientation and orbital phase. 

To investigate how these effects affect our results, we conduct a simulation that includes all these effects. We now include the physical parameters like distances, masses, and relative velocities in the simulation. We consider equal-solar-mass binaries, and their distances are sampled from 10 to 200\,pc with $dN \propto D^2dD$, where $N$ is the cumulative number of wide binaries and $D$ is the distance. We adopt a semi-major axis distribution of $dN \propto a^{-1.6}da$ \citep{El-Badry2018b} and a range from 10 to $10^5$\,AU. We consider three eccentricity distributions, $\alpha=0.5$, 1, and 1.2. After randomly sampling their binary orientation and orbital phase described in Sec.~\ref{sec:projection-effect}, we compute their $v$-$r$ angles, projected separations, projected orbital velocities, and proper motion differences. We then adopt a constant $\sigma_{\Delta \mu}=0.1$\masyr, which is the median value of the {\it Gaia} EDR3 wide binary catalog, and then compute $\sigma_\gamma$ using Eq.~\ref{eq:gamma-error}. Then we discard binaries that have $\Delta \mu / \sigma_{\Delta \mu} < 3$, the same criterion used in our analysis. To mimic our sample selection, binaries with projected angular separations $<1.5$\,arcsec are excluded. In the end, we apply the Bayesian approach to infer the eccentricity distribution from the $v$-$r$ angle distribution in each projected separation bin.

\begin{figure}
	\centering
	\includegraphics[width=\linewidth]{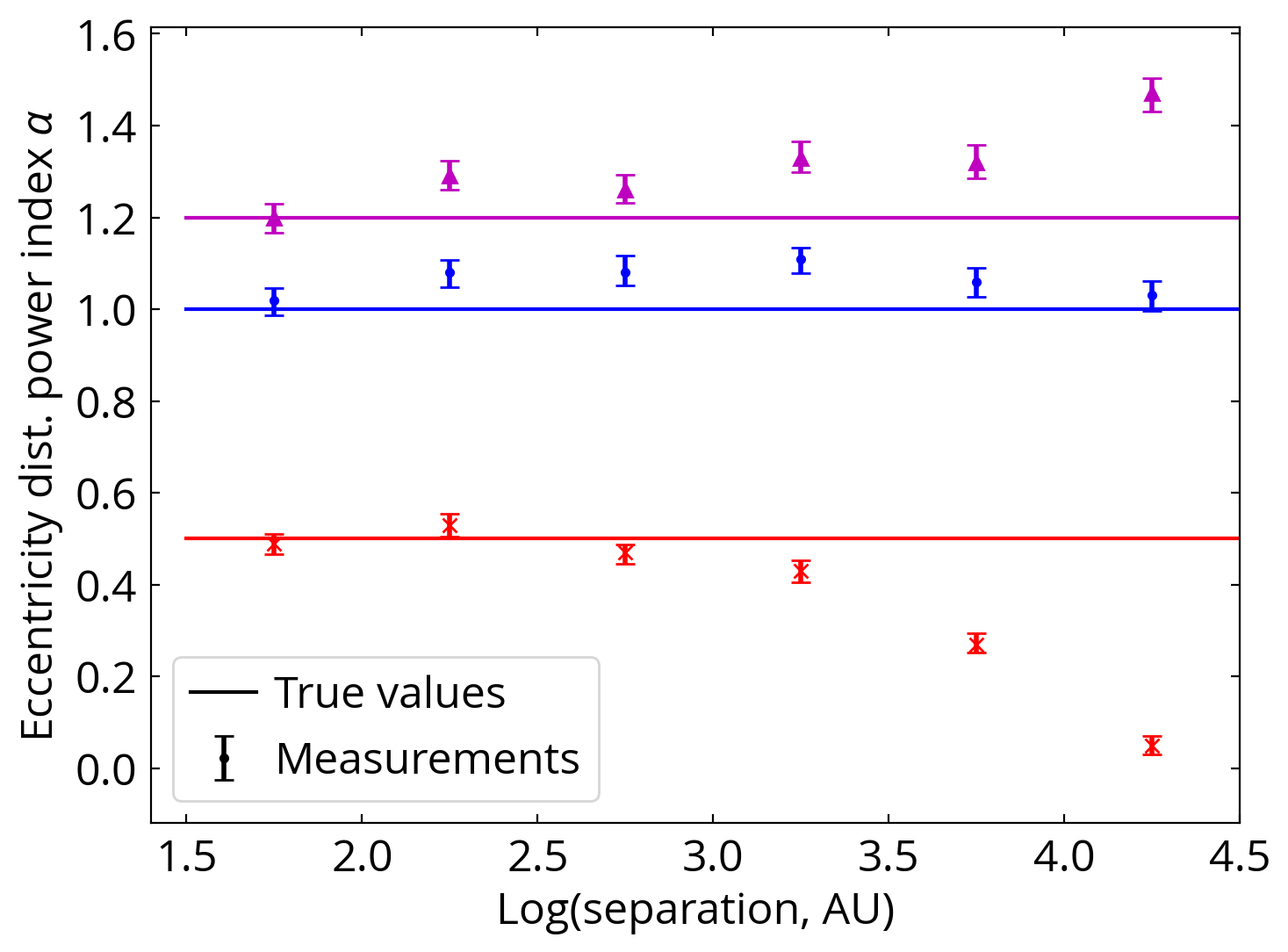}
	\caption{Test of selection effects by measuring the eccentricity distributions for simulated wide binaries. The solid lines show the input eccentricity distributions, and the markers present the measurements. Our method recovers the thermal eccentricity distribution well, but underestimates/overestimates $\alpha$ for subthermal/superthermal eccentricity distributions at $>10^3$\,AU due to the SNR criterion on the proper motion difference. This test suggests that our measured $\alpha=1.3$ at $>10^{3.5}$\,AU may be slightly overestimated from a true $\alpha \sim1.2$, but this does not change the main conclusion that the eccentricity distribution is superthermal at $>10^{3}$\,AU. }
	\label{fig:systematics}
\end{figure}

Fig.~\ref{fig:systematics} shows the simulation results. The solid horizontal lines show the true $\alpha$ values for each case. The results show that, for the thermal eccentricity $\alpha=1$, our approach can correctly recover their underlying eccentricity distribution out to $10^{4.5}$\,AU. The slight overestimate of $\alpha \sim1.1$ at $\sim10^3$\,AU is likely due to the time-averaging effect. The $v$-$r$ angle SNR criterion is the dominant selection effect for non-thermal eccentricity distributions at separations $>10^{3.5}$\,AU, making the measured $\alpha$ values deviating in the direction away from the thermal distribution.

The reason why the thermal eccentricity distribution is almost not affected by the $v$-$r$ angle SNR criterion is due to a serendipitous property of the thermal eccentricity distribution. We simulate a face-on binary sample with a thermal eccentricity distribution, with all their distances and binary separations fixed. We find that, for any values of the minimum orbital velocity we apply to the sample, the resulting $v$-$r$ angle distribution remains the same. Therefore, for binaries with the projection effects and the thermal eccentricity distribution, the resulting $v$-$r$ angle distribution is always flat regardless of the velocity (proper motion difference) criterion used.

Fig.~\ref{fig:systematics} implies that the selection effects cannot explain the strong change in $\alpha$ at $10^2$ to $10^3$\,AU in Fig.~\ref{fig:alpha-sep}. At $>10^{3.5}$\,AU, the measured $\alpha=1.3$ may be slightly overestimated from a true value of $\sim1.2$, but it does not change the main conclusion that the eccentricity distribution is superthermal at $>10^{3}$\,AU.

For a sample more distant than the 200-pc sample studied here, the SNR criterion would exclude a significant fraction of wide binaries and distort the observed $v$-$r$ angle distribution. In this case, a better approach to infer the population eccentricity distribution is probably to include the SNR distribution (e.g. the fraction of low-SNR objects is related to the underlying eccentricity distribution) in the Bayesian model (Eq.~\ref{eq:baye-e-dist}) without excluding these low-SNR wide binaries.

If a wide binary has an unresolved companion and forms a hierarchical triple, in principle our method measures the eccentricity of the outer orbit if the measured proper motions reflect the motion of the barycenter of the unresolved system. However, the presence of an unresolved companion can affect the observed proper motions by the inner orbital motion, which strongly depends on the flux ratios and therefore mass ratios \citep{Belokurov2020}. This effect tends to randomize the observed $v$-$r$ angles and produce a uniform $v$-$r$ distribution, mimicking the thermal eccentricity distribution. Although about half of the wide pairs may be hierarchical triples \citep{Moe2021binaryplanet}, we estimate that only 20-30\% of wide pairs at $10^4$\,AU have inner orbits with relevant semi-major axes and mass ratios to affect the observed proper motions, and the effect is weaker for binaries with smaller separations. Furthermore, this effect cannot explain the observed superthermal eccentricity distribution. Therefore, we do not expect the effect of unresolved companions to be important in our main results.

To summarize, in this section we discuss several possible systematics, including the contamination from chance-alignment pairs, projection effects, the SNR criterion on the proper motion difference, and the presence of unresolved companions. The most important effect is from the SNR criterion on the proper motion difference which affects non-thermal eccentricity distributions at large binary separations, but this effect does not change our main conclusions. Therefore, we consider our results robust over these possible systematics.

\section{Discussion}
\label{sec:discussion}

\subsection{Literature comparison}

Most previous studies of binary eccentricity distributions focus on close binaries at $\lesssim100$\,AU because their orbital periods are short. By using a volume-limited sample within 25\,pc, \cite{Raghavan2010} show that the eccentricity distribution of solar-type stars is close to uniform for orbital periods between 12 and $10^6$\,day ($\sim300$\,AU), and most binaries have circular orbits below 12 days due to tidal circularization \citep{Mathieu1994,Zahn2008,Price-Whelan2018}. The review by \cite{Duchene2013} further concludes that for binaries with periods of $10^2$-$10^4$\,days (roughly 0.5-10\,AU) and different masses, the eccentricity distributions are more consistent with a uniform distribution compared to the thermal distribution. \cite{Moe2017} report a similar finding of uniform eccentricity distributions for primary masses of 0.8-5\Msun\ and periods $<10^5$\,day (50\,AU), except that binaries with primary masses $>5$\Msun\ may have a thermal eccentricity distribution at periods $>10$\,day. Therefore, these studies all show that the eccentricity distribution of binaries at $<100$\,AU is consistent with a uniform distribution, in agreement with our finding.

Our results show that the eccentricity distribution is close to uniform at $100$\,AU and becomes thermal at $>10^{2.5}$-$10^3$\,AU. This agrees with the conclusions from \cite{Tokovinin2020a} where he reports that wide binaries at $10^3$-$10^4$\,AU have a nearly thermal eccentricity distribution, and those at $<200$\,AU have less eccentric orbits. Furthermore, we confirm the finding of \cite{Tokovinin2020a} that the eccentricity distribution is superthermal at $>10^3$\,AU, which can be clearly seen in the $v$-$r$ angle distribution (Fig.~\ref{fig:gaia-vrangle}).

In terms of methodology, we infer population eccentricity distributions using $v$-$r$ angles alone, and \cite{Tokovinin2020a} uses $v$-$r$ angles and additional information from the amplitude of relative velocities normalized by the expected circular velocities, which requires masses and distances of the binary. With the assumption that binary orientation is random, both methods ($v$-$r$ angle distributions versus two-dimensional angle-velocity distributions) are uniquely determined by the underlying eccentricity distributions, thus sharing identical information content. In terms of inferring individual eccentricities like Sec.~\ref{sec:e-individual}, the additional velocity information helps better constrain those eccentric binaries having observed relative velocities larger than expected circular velocities. However, for binaries with measured relative velocities smaller than circular velocities, the additional velocity information does not improve the individual eccentricity inference. This is because relative velocities higher than circular velocities can only be explained by eccentric orbits near pericenters, but relative velocities lower than circular velocities (which are more common than the former case) can be caused by both binary orientation and eccentric orbits around apocenters, and $v$-$r$ angles cannot differentiate apocenters and pericenters (Fig.~\ref{fig:orbit}).

To ensure reliable mass estimates, \cite{Tokovinin2020a} excludes unresolved triples and higher-order multiples based on external catalogs, while we do not explicitly exclude them because our method does not rely on masses. These unresolved triples would have less eccentric outer orbits due to dynamical stability \citep{Shatsky2001,Tokovinin2016}. Therefore, these triples would contribute more less-eccentric orbits in our sample compared to \cite{Tokovinin2020a}. To test this scenario, we select two samples, one with brighter main-sequence primaries ($-1<\Delta G<-0.4$) and one with fainter main-sequence primaries ($-0.2<\Delta G<0.3$), where $\Delta G$ is the absolute $G$-band magnitudes offset from the Pleiades main sequence and negative $\Delta G$ means brighter than the Pleiades main sequence \citep{Hamer2019, Hwang2020b}. We further require the primaries to have $BP$-$RP$ colors between 0.5 and 2\,mag, binary separations between 10$^2$ and $10^3$\,AU, and other criteria following Sec.~\ref{sec:gaia-vr}. We find that the brighter sample, which has more unresolved companions, has a less eccentric eccentricity distribution ($\alpha=0.75^{+0.11}_{-0.11}$) than the fainter sample ($\alpha=1.01^{+0.05}_{-0.05}$). This agrees with the picture that the wide companions of triples are less eccentric than the wide binaries without subsystems, although future investigation is needed to control other differences between the two samples. This result also suggests that unresolved companions do not strongly affect our $v$-$r$ angle measurements of wide companions; otherwise, if the unresolved companions strongly induce noise in measured $v$-$r$ angles, we would expect a more uniform $v$-$r$ angle distribution for the brighter sample.

Therefore, our results agree well with the literature that the eccentricity distribution is uniform at $<100$\,AU and gradually becomes superthermal at larger separations. In addition to using the largest wide binary sample to date, our Bayesian approach robustly incorporates the measurement uncertainties of $v$-$r$ angles and provides realistic uncertainties for the resulting eccentricity distribution measurements.

\subsection{Implications for binary formation}

\cite{Jeans1919} first showed that when a population of binaries reaches a thermal equilibrium where the energy distribution follows a Boltzmann distribution, their eccentricity distribution is $f_e(e)de=2ede$, independent of binary separations. \cite{Ambartsumian1937} proved that the eccentricity distribution is $f_e(e)de=2ede$ when the distribution function only depends on the energy, and therefore the energy distribution does not necessarily need to follow a Boltzmann distribution to have a thermal distribution in eccentricity. Later studies further showed that the distribution of semi-major axes of the binary population does not have an equilibrium state, and under gravitational interactions, soft binaries become softer and hard binaries become harder \citep{Heggie1975}. In contrast to binary separations, the eccentricity distribution does have a steady state, and after sufficient dynamical interactions, the eccentricity distribution tends toward $f_e(e)de=2ede$ (e.g. \citealt{Geller2019}). Therefore, in the following discussion, we refer to a ``thermalization of binaries'' as the process in which the eccentricity distribution tends toward a thermal distribution.

It is unlikely that the thermal eccentricity distribution at $10^3$\,AU and superthermal at $>10^3$\,AU is due to the gravitational interaction with nearby passing stars and molecular clouds. First, binaries with separations of $10^3$\,AU have a timescale longer than a Hubble time for external gravitational interactions to be dynamically important \citep{Heggie1975,Weinberg1987}. Second, the gravitational interaction tends to thermalize the wide binaries and cannot explain the observed superthermal eccentricity distribution. Therefore, given that the post-formation interaction plays a minor role, the eccentricity distribution at $<10^4$\,AU is mostly imprinted by binary formation.

Wide binaries at $>100$\,AU can form from turbulent fragmentation \citep{Bate2009,Bate2014,Offner2010}, dissolution of clusters \citep{Kouwenhoven2010,Moeckel2011}, dynamical unfolding of compact triples \citep{Reipurth2012}, and random pairings of pre-stellar cores \citep{Tokovinin2017}. Different formation channels result in different eccentricity distributions. In the scenario of the dynamical unfolding of compact triples, the eccentricity of the outer obits at $>1000$\,AU increases with increasing semi-major axis, and the majority of outer orbits at $>10^4$\,AU have $e>0.9$ \citep{Reipurth2012}. The eccentricity distribution for binaries at $>10^3$\,AU resulting from the dissolution of clusters is close to thermal \citep{Kouwenhoven2010}. For turbulent fragmentation, the radiation hydrodynamical simulations show that nearly all wide binaries at $>10^3$\,AU have $e>0.6$ \citep{Bate2014}. Thus, an observed superthermal eccentricity distribution suggests that the dissolution of clusters alone is insufficient and another process which tends to produce eccentric orbits (e.g. turbulent fragmentation and/or dynamical unfolding) must also contribute.

The observed sizes of protoplanetary disks are $\sim100$\,AU \citep{Andrews2018disk, Huang2018}, and close binaries at $\lesssim100$\,AU may form from the gravitational instability of the disks (disk fragmentation, \citealt{Kratter2006,Moe2019, Tokovinin2020}). After its birth in the disk, the binary can be initially eccentric due to the $m=1$ mode perturbation \citep{Shu1990,Krumholz2007, Kratter2010, Kratter2011a}, surrounded by the circumbinary disk. The interaction between the binary and the circumbinary disk induces orbital migration and excites the binary eccentricity \citep{Artymowicz1991,Artymowicz1994}, and the effect depends on the properties of the disk and the binary \citep{Pichardo2005,Ragusa2020,Heath2020}. Due to the complicated binary-disk interaction, the resulting eccentricity distribution from close binary formation remains unclear. Forming low-mass stars (K- and M-dwarfs) from disk fragmentation may be difficult because their disk is not sufficiently massive for gravitational instability \citep{Kratter2008}. Instead, such binaries may be formed from the dynamical interaction of unstable multiples \citep{Bate2002binary}, and their eccentricity distribution is subject to the interaction with disk and gas. 

It is noteworthy that the separation range $10^{2}$-$10^{3}$\,AU, where the eccentricity distribution changes from a uniform to a thermal one, coincides with the range of separations where the binary fraction dependence on metallicity changes as well. Specifically, below $100$\,AU, the binary fraction is anti-correlated with the metallicity \citep{Raghavan2010,Badenes2018,Moe2019,Mazzola2020}. Such anti-correlation with metallicity disappear at $\sim200$\,AU \citep{El-Badry2019a}, and the wide binary fraction at $10^3$-$10^4$\,AU has a non-monotonic relation with the metallicity \citep{Hwang2021a}. These results suggest that two different formation mechanisms are operating at binary separations above and below $\sim1000$\,AU. Another important property of wide binaries that changes in the same range of separations is mass ratios. Equal-mass binaries are common in close binaries, whereas wide binaries appear to be independently drawn from the initial mass function \citep{Moe2017}. But at intermediate separations up to $\sim 1000$ AU, \cite{El-Badry2019} find an enhancement of equal-mass binaries, consistent with both formation scenarios contributing to the intermediate regime of $10^2$-$10^3$\,AU separations.

Similarly, we interpret the observed change in the eccentricity distribution as resulting from different binary formation mechanisms dominating at different separations. The close binary formation at $<10^{2}$\,AU, most likely from the disk fragmentation \citep{Kratter2006} and further interactions of the binary with disk, results in a uniform eccentricity distribution. Why disk fragmentation results in a uniform eccentricity distribution remains unknown, and future studies are needed to investigate this connection. The wide binary formation dominating at $>10^{3}$\,AU leads to a superthermal eccentricity distribution. Both close and wide binary formations contribute to the transition separation of $10^{2}$-$10^{3}$\,AU.

The superthermal eccentricity distribution at $>10^3$\,AU suggests that the cluster dissolution scenario cannot be the only wide-binary formation channel, which predicts a thermal eccentricity distribution \citep{Kouwenhoven2010}. The superthermal eccentricity distribution requires highly eccentric binaries at $\sim10^3$\,AU, consistent with the turbulent fragmentation \citep{Bate2014} and the dynamical unfolding of compact triples \citep{Reipurth2012}, where both scenarios only have eccentric binaries at $>10^3$\,AU. Both scenarios give extremely eccentric ($e>0.9$) binaries at $\gtrsim10^4$\,AU and a strong positive correlation between binary separation and eccentricity at $>10^3$\,AU, but our results show that $\alpha$ remains roughly constant from $10^3$ up to $10^{4.5}$\,AU. Our result has a large error bar at $>10^4$\,AU and may be affected by the systematics from the SNR criterion on proper motion differences and unresolved companions (Sec.~\ref{sec:systematics}). Furthermore, at $>10^4$\,AU, gravitational interactions with passing stars, molecular clouds, and the Galactic tide become important \citep{Weinberg1987,Jiang2010a}, which may alter the eccentricity distribution.

The definition of close and wide binaries are often arbitrary. The eccentricity distribution presented here shows that close binaries dominate at $<10^{2}$\,AU and wide binaries at $>10^{3}$\,AU, with transition in between. Therefore, the eccentricity distribution provides us a clean distinction for what close and wide binaries are. In this work, we demonstrate that in addition to the metallicity dependence and mass-ratio distribution, eccentricity is emerging as a powerful tool to unravel the formation channels of binaries.

\subsection{Recommendation for the use of the catalog}

In the catalog, we provide eccentricity measurements of wide binaries using the separation-dependent eccentricity prior (Eq.~\ref{eq:alpha-best-fit}). While we compute eccentricities for all wide binaries in \cite{El-Badry2021}, the user should make sure that the wide binary is not a chance alignment (e.g. use the chance-alignment probability from \citealt{El-Badry2021}, or acquire radial velocity data) and make sure that {\it Gaia}'s astrometric measurements are robust (e.g. use astrometric quality indicators from Gaia). A significant non-zero proper motion difference (e.g. \texttt{dpm\_sig}$>3$) is needed for $v$-$r$ angle measurements and the eccentricity inference. If the user has a high-confidence wide binary that is not included in our catalog, the user can compute its eccentricity posterior using our Bayesian framework, and the simulated grid of likelihood is available online \footnote{\url{https://github.com/HC-Hwang/Eccentricity-of-wide-binaries}}.

Our approach only uses parallaxes (distances) of wide binaries for the binary separation computation and thus for the prior $p(e)$. If the user has a wide binary without reliable parallaxes, one can still compute its eccentricity posterior either by adopting an assumed prior or by some binary separation estimate (e.g. photometric distance), which is often sufficient for an estimated eccentricity distribution from Eq.~\ref{eq:alpha-best-fit} or Fig.~\ref{fig:alpha-sep}.

{\it Gaia} systematics affects pairs with angular separations $<1$\,arcsec (Sec.~\ref{sec:gaia-systematics}), and thus their inferred eccentricity posteriors are not robust. Therefore, we do not recommend using the eccentricity measurements for pairs at $<1$\,arcsec. The distribution of the most probable eccentricities may have some artifacts because the eccentricity errors depend on eccentricities (Fig.~\ref{fig:e-pdf}). Therefore, we do not recommend using the most probable eccentricities only to conduct population study.

The assumptions used in our eccentricity inference is that the two stars are in a Keplerian orbit and that their orientation and orbital phase are random. The assumption of Keplerian orbits still holds for hierarchical triples because the dynamics of the outer companion can be well described by a Keplerian orbit when the proper motion measurements are not strongly affected by the inner orbit. For rare dynamically unstable triples, the eccentricity inference does not apply, but the $v$-$r$ angle measurements can still provide information on their dynamics. Unbound binaries that are heading in the opposite directions would have $v$-$r$ angles close to 0$^\circ$, resulting in an (incorrect) inferred eccentricity close to 1. Fig.~\ref{fig:gaia-vrangle} shows that the observed $v$-$r$ angle distributions are very symmetric, suggesting that there is no significant number of unbound binaries in the sample. Although this is expected because unbound binaries spend small amount of time at separations $<10^4$\,AU, these binaries, if they exist, are likely excluded from \cite{El-Badry2021} catalog because their relative velocity does not follow the Keplerian law. For non-Newtonian gravity, wide binaries at $\gtrsim 7000$\,AU would deviate from Keplerian orbits and thus the $v$-$r$ angle distribution can be an independent test on gravity theory \citep{Banik2018, Banik2021}. 

\section{Conclusions}
\label{sec:conclusion}

Eccentricities of wide binaries play a critical role in understanding binary formation. However, it is challenging to measure their eccentricities due to their long orbital periods. In this paper, we use the $v$-$r$ angles to measure eccentricities for wide binaries at $10^{1.5}$ to $10^{4.5}$\,AU. Our method only requires a minimal assumption and observed quantities and do not require accurate masses and distances (parallaxes). We provide an electronic catalog that includes the individual eccentricity measurements for {\it Gaia} EDR3 wide binaries. Our findings include:

\begin{enumerate}
    \item The eccentricity distribution of wide binaries is close to uniform at $\sim100$\,AU, reaches thermal ($f_e(e)=2e$) at $\sim10^{2.7}$\,AU, and becomes superthermal at $>10^{3}$\,AU (Fig.~\ref{fig:alpha-sep}). 
    \item Since the cross-section of a $10^3$\,AU binary is too small for gravitational interactions with passing stars and molecular clouds, the observed eccentricity distribution is largely imprinted by binary formation. Therefore, the eccentricity distribution provides a clear distinction between close and wide binaries, with close binaries dominating at $<10^2$\,AU and wide binaries at $>10^3$\,AU.
    \item The close binary formation, most likely disk fragmentation, dominates at $<10^{2}$\,AU, resulting in a uniform eccentricity distribution. The wide binary formation results in the superthermal eccentricity distribution at $>10^{3}$\,AU. The dissolution of natal stellar cluster acting alone would result in a thermal distribution and therefore cannot produce such high eccentricities. We conclude that other processes that results in high eccentricities -- for example, turbulent fragmentation and/or the unfolding of compact triples -- must also contribute.
\end{enumerate}

The authors are grateful to the referee for the detailed and constructive report which significantly improved the paper. The authors appreciate Andrei Tokovinin's constructive comments on the manuscript. HCH thanks Scott Tremaine for the explanation on the relation between thermal eccentricity distribution and the uniform observed $v$-$r$ angles. HCH appreciates discussions with Canon Sun, Aaron Geller, Chris Hamilton, and Vedant Chandra. HCH acknowledges the support of the Infosys Membership at the Institute for Advanced Study and from Space@Hopkins. YST acknowledges financial support from the Australian Research Council through DECRA Fellowship DE220101520. NLZ acknowledges the support of the J. Robert Oppenheimer Visiting Professorship and the Bershadsky Fund at the Institute for Advanced Study.

{\it Facilities:} Gaia.

{\it Software:} \texttt{IPython} \citep{ipython2007}, \texttt{jupyter} \citep{jupyter2016}, \texttt{Astropy} \citep{Astropy2013,Astropy2018}, \texttt{numpy} \citep{numpy2020}, \texttt{scipy} \citep{scipy2020}, \texttt{matplotlib} \citep{matplotlib2007}, \texttt{Mathematica} \citep{mathematica}, \texttt{emcee} \citep{Foreman-Mackey2013}.

\section*{Data Availability}
The data underlying this article are available in the article and in its online supplementary material.

\appendix

\section{Formal expressions for \lowercase{$p(\gamma|e)$} and \lowercase{$p(\gamma|\alpha)$} including projection effects}
\label{sec:appendix}

The task of this appendix is to obtain a general expression for the distribution function of projected $v$-$r$ angles at a particular value of eccentricity, $p(\gamma|e)$, given random sampling of the orbits, random orientations of the orbits relative to the observer, and geometric projection effects. 
In the main paper, we use a numerical simulation with $10^6$ binaries appropriately sampling all parameters to obtain $p(\gamma|e)$ and $p(\gamma|\alpha)$. These distributions are shown in Fig. \ref{fig:vr-e}, right and Fig. \ref{fig:gamma-bar-e}, right.  

We use Eq. \ref{eq:vr-angle-xy} where we express $\vec{v}_{XY}$ and $\vec{r}_{XY}$ as a function of orbital and orientation parameters to obtain the following explicit expression for $\cos \gamma$:
\begin{equation}
\label{eq:gamma-XY-ana}
    \cos \gamma = \frac{e \cos^2\iota \cos \omega \sin(f+\omega) - \cos(f+\omega)(e\sin \omega + \sin^2 \iota \sin(f+\omega))}
    {\sqrt{\cos^2(f+\omega) + \cos^2\iota \sin^2(f+\omega)} \sqrt{\cos^2 \iota (e \cos \omega + \cos(f+\omega))^2+ (e\sin \omega + \sin(f+\omega))^2}}
\end{equation}
Here inclination $\iota$ is defined to be the angle between the orbital plane and the plane of the sky (or equivalently, the angle between the angular momentum vector and the line of sight). For randomly oriented orbital planes, the angular momentum vector is uniform on a sphere, and the probability density function of $\iota$ is $(\sin\iota)/2$ defined on $\iota\in[0,\pi]$. The distribution of $\omega$ is uniform on $[0,2\pi]$. The distribution of true anomaly $f$ is given by Eq. \ref{eq:p-f}. 

Now the task is to find the distribution of $\gamma$ if the distributions of all orbital and geometric parameters it depends on via Eq. \ref{eq:gamma-XY-ana} are known. Using the Dirac delta function for changing variables in a multi-variate probability distribution function and the single-variable change $p(\gamma)=p(\cos\gamma)\sin\gamma$, we formally obtain
\begin{equation}
p(\gamma|e)=\sin \gamma \int_0^{2\pi} \frac{{\rm d}\omega}{2\pi} \int_0^{\pi} \frac{\sin\iota {\rm d}\iota }{2} \int_0^{2\pi} \frac{{\rm d} f}{2\pi}  \delta\left(  \cos \gamma - {\rm RHS}(e,\omega,\iota,f)\right)\frac{(1-e^2)^{3/2}}{(1+e\cos f)^2}.
\label{eq:app_gamma_e}
\end{equation}
Here ${\rm RHS}(e,\omega,\iota,f)$ is the right hand side of Eq. \ref{eq:gamma-XY-ana}. 

For a given distribution of eccentricities $f_e(e)$, $p(\gamma)=\int p(\gamma|e)f_e(e){\rm d}e$, and in particular for the power-law distributions we consider in this paper,  
\begin{equation}
p(\gamma|\alpha)=\int_0^1 p(\gamma|e)(\alpha+1)e^{\alpha}{\rm d}e.
\label{eq:app_gamma_alpha}
\end{equation}

Eq. \ref{eq:app_gamma_e} and \ref{eq:app_gamma_alpha} can be numerically evaluated. To this end, the Dirac delta function can be approximated by an appropriately normalized Gaussian. Then the multi-dimensional integrals can be numerically evaluated with a variety of methods (e.g., via Monte-Carlo sampling). Narrower Gaussians result in a better approximation but require a finer sampling of the integration domain. We have confirmed that the numerical evaluations of Eq. \ref{eq:app_gamma_e} and \ref{eq:app_gamma_alpha} yield the same result as the simulations shown in Fig. \ref{fig:vr-e} and \ref{fig:gamma-bar-e}.



\section{Generalized eccentricity distribution}
\label{sec:generalized-e}

Here we lay out a procedure for inferring a generalized eccentricity distribution $p(e|\{g_k\})$, where $\{g_k\}$ is a set of free parameters to describe the eccentricity distribution. Similar to Eq.~\ref{eq:baye-e-dist}, the best fit of the parameters for a given observed $v$-$r$ angle distribution $\{ \gamma_{obs,i} \}$ can be determined by
\begin{equation}
\label{eq:baye-generalized}
\begin{multlined}
p(\{g_k\} | \{\gamma_{obs,i}\}) \propto \\ \Pi_i \int p(\gamma_{obs,i} | \gamma_{true,i}) p(\gamma_{true,i}| e_i)  p(e_i|\{g_k\}) d\gamma_{true,i} de_i,
\end{multlined}
\end{equation}
where $i$ is the index of individual wide binaries. In the main text, we use a power law to describe the eccentricity distribution (Eq.~\ref{eq:p-e-bar-alpha}). Since there is only one free parameter $\alpha$ in the power law, it is straightforward to explore the parameter space and obtain the best fit and corresponding uncertainties. However, when the number of free parameters is more than one, exploring the entire parameter space becomes challenging, and it is more feasible to obtain the posterior distributions using the Markov-Chain Monte-Carlo (MCMC) method.

As an example, we consider a multi-step function for the eccentricity distribution such that 
\begin{equation}
\label{eq:step-func}
\begin{multlined}
p(e|\{g_k\})=g_k{\rm\ for\ } e_k\le e<e_{k+1}.
\end{multlined}
\end{equation}
Its normalization requires that $\Delta e \sum g_k  = 1$, where $\Delta e = e_{k+1} - e_k$ for equally spaced eccentricity bins. In principle, any distribution is describable by this multi-step function with sufficiently narrow $e_k$ bins. Assuming that (1) $\{ \gamma_{obs,i} \}$ measurement uncertainties are negligible and (2) the eccentricity bin $\Delta e$ is sufficiently small so that $p(\gamma|e)$ does not vary much within a bin, we can simplify Eq.~\ref{eq:baye-generalized} as 
\begin{equation}
\label{eq:baye-multi-step-simplified}
\begin{multlined}
p(\{g_k\} | \{\gamma_{obs,i}\}) \propto \\ \Pi_i  \left( \sum_k p(\gamma_{obs,i}|e_k') g_k \right),
\end{multlined}
\end{equation}
where $e_k'=(e_k+e_{k+1})/2$, the center of the bin. 

With the simplified probability Eq.~\ref{eq:baye-multi-step-simplified}, we use \texttt{emcee} \citep{Foreman-Mackey2013} to sample the posterior distribution. We use an eccentricity step of 0.1, and therefore we have $10-1=9$ free parameters $g_k$ due to the normalization criterion. We choose $k=1$ to 9 as free parameters and then $g_0$ for $0<e<0.1$ is computed by $g_0=10-\sum_{k=1}^9 g_k$. We use priors that $g_k>0$ and $p(e|\{g_k\})$ is properly normalized (i.e. $\sum_{k=1}^9 g_k\le10$). We apply this analysis to the wide binaries with separations between $10^3$ and $10^{3.5}$\,AU, with other selections identical to the main text. 

\begin{figure}
	\centering
	\includegraphics[width=0.8\linewidth]{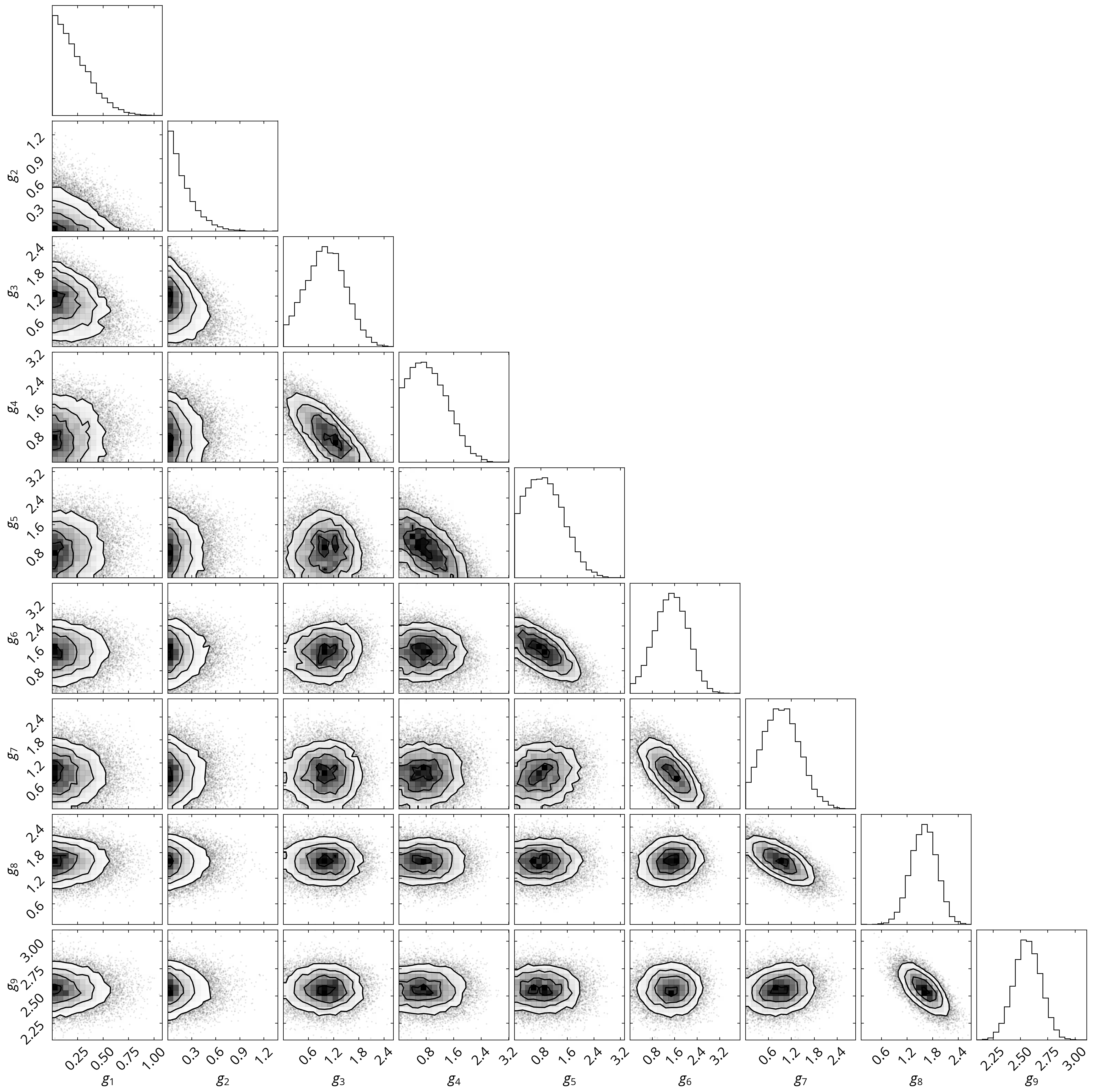}
	\caption{Posterior distributions for the eccentricity distribution at $10^3$-$10^{3.5}$\,AU, formulated by a multi-step function. }
	\label{fig:corner}
\end{figure}

\begin{figure}
	\centering
	\includegraphics[width=0.5\linewidth]{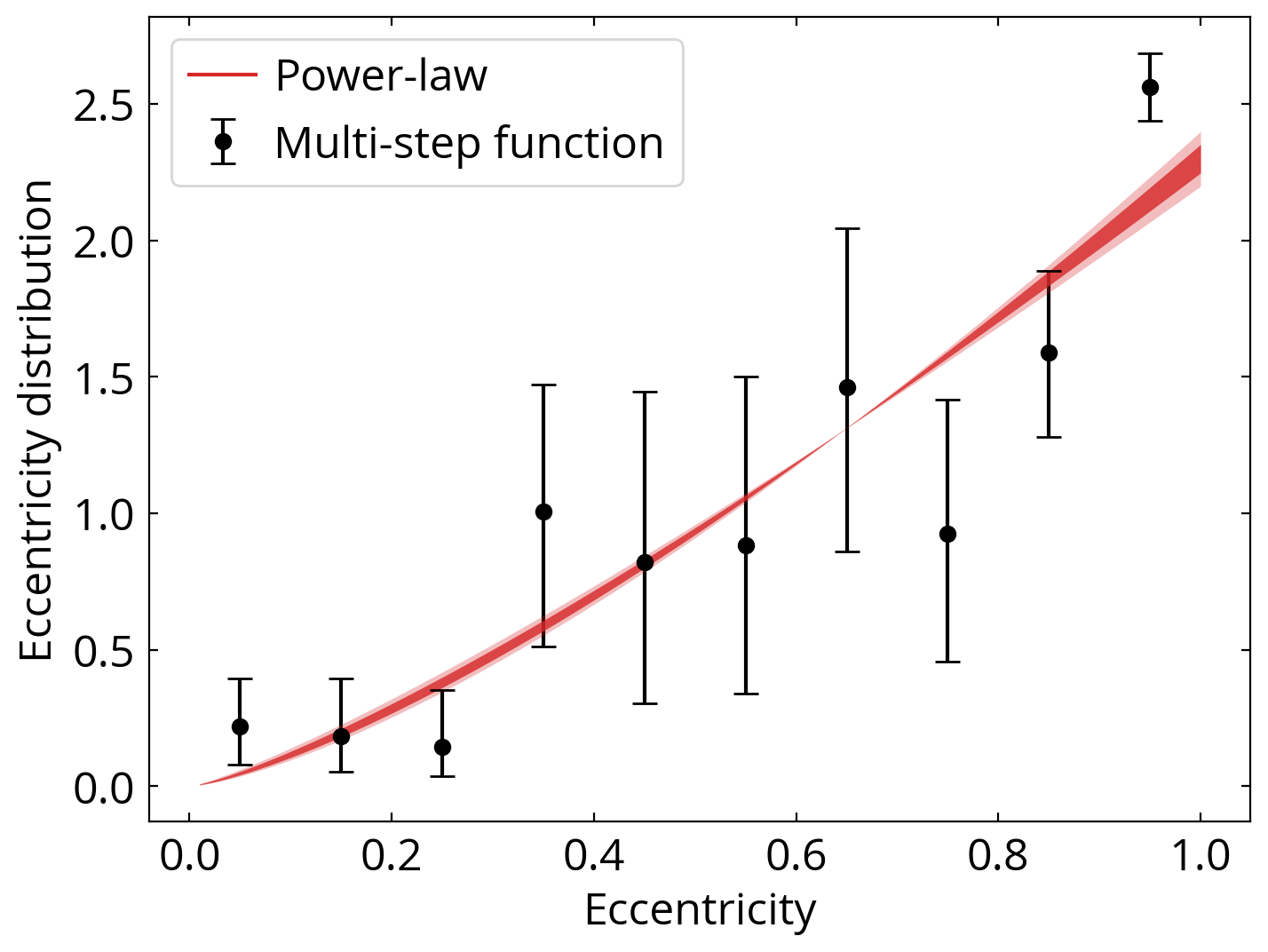}
	\caption{Comparison between a power-law formulation (red line, Sec.~\ref{sec:e-dist}) and a multi-step function formulation (black points, Appendix~\ref{sec:generalized-e}) for the eccentricity distribution of wide binaries at $10^3$-$10^{3.5}$\,AU. Both formulations agree well that the low-eccentricity ($e<0.3$) wide binaries are suppressed and the highly eccentric ($e>0.9$) binaries are enhanced, making the eccentricity distribution super-thermal. The free parameters at $0.3<e<0.8$ in the multi-step function formulation suffer from strong degeneracy and thus have larger uncertainties. }
	\label{fig:multi-step-e}
\end{figure}

Fig.~\ref{fig:corner} shows the resulting posterior distributions made using \texttt{corner} \citep{Foreman-Mackey2016}. When using a muti-step function to parametrize the eccentricity distribution (which is also used in \citealt{Tokovinin2020a}), the adjacent values ($g_k$ and $g_{k+1}$) are often strongly degenerate because the adjacent eccentricities have similar $p(\gamma|e)$. Fig.~\ref{fig:multi-step-e} shows the measured eccentricity distribution, with the error bars indicating the 16-50-84 percentiles. We overplot the result from the power-law description obtained in Sec.~\ref{sec:e-dist} and Fig.~\ref{fig:e-dist}. The power-law result (red line) agrees well with the multi-step function result (black points), suggesting that our choice of using a power-law to parametrize the eccentricity distribution is reasonable. The result from multi-step function further supports the finding that wide binaries with $e<0.3$ are suppressed, and those with $e>0.9$ are enhanced, making the eccentricity distribution super-thermal.

The advantage of using a power-law for the eccentricity distribution is that it only has one free parameter and exploring its parameter space is straightforward. Its disadvantage is that a power-law is not suitable for exotic cases, for example single-valued eccentricity distributions, although in this case it would be obvious from its $v$-$r$ angle distributions. The advantage of using a multi-step function for the eccentricity distribution is that it is more intuitive and more generalized. For instance, we verify that it can recover single-valued eccentricity distributions. The disadvantage is that it involves more free parameters with strong degeneracy. This means that it requires a larger sample size than the power-law method to have sufficient constraining power. Furthermore, it requires more advanced technique like MCMC to explore the parameter space. Here, for the purpose of demonstration, we simplify the probability in Eq.~\ref{eq:baye-multi-step-simplified} by neglecting the measurement uncertainties, but these terms should be included for a formal analysis, which may slow down the MCMC. Alternatively, one can choose other functional form for the eccentricity distribution where the free parameters are not strongly degenerate, for example the beta distribution \citep{Hogg2010a}.

A related question is whether the eccentricity distribution and the $v$-$r$ angle distribution have a unique mapping onto each other, or whether it is possible that two different eccentricity distributions can result in the same $v$-$r$ angle distribution. In Fig.~\ref{fig:vr-e} (right), we numerically show that $v$-$r$ angle distributions are different for different power-law indices when the eccentricity distribution is a power-law. Therefore, this is sufficient for this study where the observed eccentricity distribution seems to agree with the power law. We suspect this remains approximately true for other eccentricity distributions. However, there is no one-to-one relation if the orientation is not random (e.g. transiting planet hosts, \citealt{Behmard2022}). One special case is that both highly eccentric ($e\sim1$) binaries (whatever the orientation is) and edge-on binaries (whatever the eccentricity is) have the observed $v$-$r$ angle distributions peaking at $0^\circ$ and $180^\circ$ and zero elsewhere.  

To summarize, here we provide a procedure for generalized eccentricity distributions. By comparing with the results from multi-step function, we find our power-law descriptions agree well with the observations. 

\bibliography{wide_binary_eccentricity}{}
\bibliographystyle{aasjournal}
\end{document}